\documentclass[prc,aps,tighten,preprint,showpacs]{revtex4}

\usepackage{graphicx}% Include figure files
\usepackage{dcolumn}% Align table columns on decimal point
\usepackage{bm}% bold math
\usepackage{hyperref}
\usepackage{amsmath,amssymb}
\usepackage{float}
\usepackage{subfig}
\usepackage{slashed}
\usepackage{caption}
\usepackage{color}
\usepackage{grffile}
\usepackage{fancybox}
\allowdisplaybreaks

\begin{document}

\title{Hard photodisintegration of $^3\text{He}$ into a pd pair}
\author{ 
 Dhiraj Maheswari %{\thanks{dmahe001@fiu.edu}}
 and Misak M. Sargsian %\thanks{sargsian@fiu.edu} 
 }
\affiliation{Florida International University \\ Miami, FL, 33199, USA}
\date{\today}

\begin{abstract}
The recent measurements of  high energy photodisintegration of a $^3\text{He}$ nucleus to a $pd$ pair at $90^0$ center of mass demonstrated an
energy scaling consistent with the  quark counting rule with an unprecedentedly large exponent of $s^{-17}$. To understand the underlying mechanism of this 
process, we extended the theoretical formalism of the hard rescattering mechanism (HRM) to  calculate  the $\gamma$ $^3$He$\rightarrow pd$ reaction.
In HRM the incoming high energy photon strikes a quark from one of the nucleons in the target which subsequently undergoes hard 
rescattering with the quarks from the other nucleons, generating a hard two-body system in the final state of the reaction.  Within the HRM we derived 
the parameter-free expression for the differential cross section of the reaction,  which is expressed through the $^3$He$\rightarrow pd$ transition spectral function, the cross section of hard $pd\rightarrow pd$ scattering, and the effective charge of the quarks being interchanged during the hard rescattering process. The numerical
estimates of all these factors resulted in the magnitude of the  cross section, which is  surprisingly in good  agreement with the data.

\end{abstract}

\pacs{24.85.+p, 25.10.+s, 25.20.-x}

\maketitle

\section{Introduction}

The large momentum transfer photoproduction reactions with two-body breakup of the nucleus represent one of the testing grounds for nuclear quantum chromodynamics (QCD). The striking characteristics of these processes is the enormous value of \textit{invariant} energy produced even at moderate incident beam energy.
The invariant energy of the photoproduction reaction is 
$s= m_T^2 + 2E_{\gamma}m_T$, which shows that  it grows for the nuclear target $A$ times faster than that of the  proton target, where $m_T$ is the mass of the target and $E_\gamma$ is the incident photon energy. Considering large and fixed center-of-mass~(cm) angles in two-body break-up  reactions allows us to provide large  momentum transfers 
$t \sim -\frac{s}{2}(1 - cos\theta_{cm})$, thus satisfying conditions for hard QCD scattering. 

Hard nuclear scattering,  in which the energy-momentum transferred to the nucleus is much larger than the nucleon masses, is one of the best  processes to probe quark degrees of freedom in the nucleus. In the hard scattering kinematic regime, we expect that only the minimal Fock components dominate in the wave function of the particles involved in the scattering. This expectation results in the prediction of  the constituent (or quark) counting rule, according to which the energy dependence of two-body hard reaction is defined by the number of fundamental constituents participating in the reaction\cite{BF73,MMT72}. 

If we consider a reaction of the type $ a + b \rightarrow c+ d$,   according to  constituent counting rule, the energy 
dependence of the hard process  should scale as
\begin{equation} \label{scaling}
\frac{d\sigma^{(ab\rightarrow cd)}}{dt} \sim	\frac{1}{s^{n_a + n_b + n_c + n_d -2}},
\end{equation}
where $n_i, i = a, b,c,d$ represent the numbers of the fundamental fields associated with respective particles involved in the process. 
For example, if $a$ is a proton, $n_a$ will equal 3, and if it is a photon, $n_a$ would be 1. 

Even though  the energy dependencies (or scaling relations) of Eq. (\ref{scaling})  do not imply the 
onset of the perturbative QCD regime, they indicate that the resolution of the probe is such that it allows us to identify the constituents 
of the hadrons that participate in the hard scattering.
In 1976 it was suggested\cite{BC76} to use the concept of the quark-counting rule to explore the QCD degrees of freedom in nuclei.  One of the best candidate reactions  was hard photodisintegration of the deuteron, $\gamma + d \rightarrow p +n$, which, according to Eq. (\ref{scaling}), should scale as $d\sigma/dt \sim s^{-11}$.  
The first such experiments  being carried out at SLAC\cite{Napolitano:1988uu,Freedman:1993nt,Belz:1995ge} and 
Jefferson Lab\cite{Bochna:1998ca,Schulte01se,Schulte:2002tx,Mirazita:2004rb,Rossi:2004qm} revealed  $s^{-11}$ scaling for photon energies 
already at $E_\gamma \ge 1$ GeV and  $\theta_{cm}=90^o$.
It is worth mentioning that the calculations based on a conventional mesonic picture of strong interaction failed to explain the observed energy scaling, which can be considered another indication that the quark degrees of freedom need to be included for an adequate description of the reaction. 
The deuteron two-body hard  photodisintegration reactions have been  used also to measure the  polarization 
observables\cite{Wijesooriya:2001yu,Adamian:2000qc,Jiang:2007ge,Zachariou:2015clw}, which were in general agreement with the quark-constituent  picture of 
hard scattering.

To check the universality of the constituent counting rule for other hard breakup reactions, the two-body reactions were extended to a $^3$He target, in which case two fast outgoing protons and slow neutron were detected in the $ \gamma + ^3$He$\rightarrow (pp) + n$ reactiont\cite{gheppn2004}.
The results of this experiment\cite{Pomerantz:2009sb} were consistent  with the $s^{-11}$ scaling in the two-proton 
hard beak up channel, but at much larger  photon energies ($E_{\gamma} > 2$ GeV) than  in the case of $pn$ breakup.  
Recently the hard two-body breakup reaction was  measured for the more complex $\gamma +$ $^3$He$\rightarrow p+d$ channel\cite{Pomerantz13}. 
According to Eq. (\ref{scaling}) such a reaction in the hard scattering regime should scale as $s^{-17}$, and surprisingly the experiment observed a scaling consistent with the exponent of 17, an unprecedented large number to be observed in two-body hard processes. 
 
In the present work, we extend the theoretical framework referred as hard rescattering mechanism(HRM) to calculate the cross section of above mentioned $\gamma ^3$He$\rightarrow pd$ reaction. The HRM model was originally developed for calculation of $\gamma d\rightarrow pn$ reactions\cite{gdpn}. The model was successful not only in verifying the $s^{-11}$ dependence but also reproducing  the  absolute magnitude of the $\gamma +d \rightarrow pn$ cross sections without free parameters 
at $\gtrsim 1$~GeV incoming photon energies  and large center of mass angles\cite{gdpn,gdpn2,gdpnang} . 
The HRM model allowed  also the calculation of polarization observables for the $\gamma d\rightarrow pn$ reaction\cite{gpdnpol},  and its prediction for the large magnitude of transferred polarization was confirmed by  the experiment of Ref.\cite{Jiang:2007ge}.
 Subsequently the HRM model was applied to the $\gamma + ^3$He $\rightarrow pp + n$ reactions\cite{gheppn}, in which two protons were produced  in the hard break-up process while the neutron was soft.  The model described the scaling properties and the cross section reasonably well and was able to explain the observed smaller cross section as compared to the deuteron break-up reaction.  In Ref.\cite{gdBB} it was shown also that HRM model can be extended to the hard break-up of the nucleus to any two-baryonic state which can be produced from the $NN$ scattering through the quark-interchange 
 interaction.   In the HRM model, a quark of the one nucleon knocked 
 out by the incoming photon rescatters with a quark of  the other nucleon leading to a production of two nucleons with large relative momentum. 
 We assume in HRM that  the quark interchange  is the dominant mechanism for the hard rescattering of two outgoing energetic nucleons. The latter assumption 
 is essential for factorization of the hard scattering kernel from the soft incalculable part of the scattering amplitude.
  
 In the present work we apply a similar rescattering  scenario for the hard break-up of a $^3$He nucleus to a $pd$ pair. Our main goal is to  check whether the HRM approach, which explicitly accounts for the quark degrees of freedom, 
will allow us to reproduce the energy and angular dependencies of the measured cross sections. 
The  article is organized as follows: Sec. II describes the kinematics and the reference frame of  the two-body break-up reaction.  In Sec. III, 
we develop the hard rescattering model for the $\gamma + ^3$He$\rightarrow p+d$  reaction, discussing in detail the nuclear amplitude which 
according to HRM provides the main contribution to the hard break-up cross section.
In Sec. IV we complete the derivation by calculating the cross section and considering the methods of estimation of  nuclear and $pd\rightarrow pd $ rescattering parts  
entering in the cross section. Section IV presents  numerical estimates and a comparison with the results of the recent experiments at $\theta_{cm} = 90^0$. 
It also gives predictions for the angular distribution of the cross section as well as energy dependencies for other $\theta_{cm}$.
Section V summarizes our results. In Appendix A, we present the details of the derivation discussed in the Sec. III. 
The discussion of the hard elastic $pd \rightarrow pd$ scattering is presented in Appendix B. Appendix C discusses the relationship between the light-front and non-relativistic $^3$He to deuteron transition wave functions.

\section{Kinematics of the Process and the Reference Frame}
\label{sectionkin}
We are considering the following two-body photodisintegration reaction:
\begin{equation} \label{reaction}
\gamma + ^3\text{He} \rightarrow p + d,
\end{equation}
where the proton and deuteron are produced at large angles  measured in the center-of-mass  reference frame of the  reaction.  
The invariant energy $s$  and momentum transfer $t$ of the reaction are defined as
\begin{eqnarray}
s & = &  (q + p_{^3\text{He}})^2  = m_{^3\text{He}}^2 + 2 q \cdot p_{^3\text{He}} = m_{^3\text{He}}^2 + 2 E_\gamma m_{^3\text{He}}  = (E^{cm}_\gamma + E^{cm}_{^3\text{He}})^2\nonumber \\
t & =  & (q - p_p)^2 = m_p^2 - 2 q\cdot p_p = m_p^2 - 2E_\gamma^{cm}(E_p^{cm} - p_p^{cm}\cos \theta_{cm}),
 \label{mandelstams}
\end{eqnarray} 
where $m_p$ and $m_{^3\text{He}}$ are masses of 
the proton and $^3$He target, respectively, and $E_\gamma$ is the incoming photon energy in the laboratory system.
The four-vectors  $q$, $p_{^3\text{He}}$, and $p_{p}$ define  the four-momenta of photon, $^3$He, and proton respectively. 
In the right hand side of Eq. (\ref{mandelstams}), we expressed $s$ and $t$ through the center-of-mass energies, momenta,  and scattering angles of interacting particles defined as
\begin{eqnarray} 
E_\gamma^{cm} & = &  \frac{1}{2\sqrt{s}}\big(s - m_{^3\text{He}}^2\big)  \texttt{,} \ \ \ \ \ \ \ \  E_{^3\text{He}} = \frac{1}{2\sqrt{s}}\big(s + m_{^3\text{He}}^2\big) \nonumber \\
E_p^{cm} & = & \frac{1}{2\sqrt{s}}\Big(s + m_p^2 - m_d^2 \Big)  \texttt{,} \ \  E_d^{cm}=\frac{1}{2\sqrt{s}}\Big(s + m_d^2 - m_p^2 \Big).
\label{initialfinalenergyinCM}
\end{eqnarray}

The one interesting property of Eq. (\ref{mandelstams}),  observed in Ref.\cite{Holt}, is the possibility to generate large 
center-of-mass energy $s$ with   moderate  energy of photon beams. This is due to the fact that, in the expression of $s$, photon 
energy is multiplied by the mass of the target.  For the case of reaction (\ref{reaction}),  for example, the photon energy, $E_\gamma  = 1$~GeV will 
generate $s$ as large as that generated by a $6$~GeV/$c$ proton beam in $pp$ scattering. This property was one of the reasons why
the quark-counting scaling was observed in the $\gamma d\rightarrow pn$ reaction for photon energies as low as $1.2$ GeV at 
cm $90^0$ break-up kinematics\cite{Mirazita:2004rb,Rossi:2004qm}.

Using Eq. (\ref{initialfinalenergyinCM}) in the expression for $t$ in Eq. (\ref{mandelstams}), we obtain
\begin{equation}\label{expr_for_t}
t=m_p^2-\frac{1}{2s}(s-m_{^3\text{He}}^2) \Big[(s+m_p^2-m_d^2)-\sqrt{\left\{ s-(m_p+m_d)^2\right\} \left\{ s-(m_p - m_d)^2 \right\}} \cos \theta_{cm} \Big].
\end{equation}
It follows from the above relation  that in the high energy limit $t\sim -\frac{s}{2}(1-cos\theta_{cm})$, which indicates that at large and fixed values of 
$\theta_{cm}$ one can achieve the hard scattering regime,  $-t(-u)\gg m_N^2$,  providing large values of $s$.  For the latter, it follows from the 
expression of $s$ in Eq. (\ref{mandelstams}) that the  photon energy $E_\gamma$ is 
multiplied by $2 m_{^3\text{He}}$, because of which,  even for moderate 
value of $E_\gamma$, the high energy condition  ($s\gg m_N^2$) is  easily achieved. 
This is seen in Fig. 1(a), where the invariant momentum transfer $-t$ is presented as a function of incoming photon energy $E_{\gamma}$ at large and fixed values of $\theta_{cm}$.  As the figure shows, even at $E_\gamma \sim 1$~GeV the invariant momentum transfer $-t\sim  1$ (GeV/c)$^2$, which is sufficiently large for the reaction to be considered hard.

%inserting momentum vs Energy and -t vs Energy graphs

\begin{figure}[ht]
\centering
\subfloat [\label{FigSub1}]{%
\includegraphics[height = 6 cm, width = 8 cm]{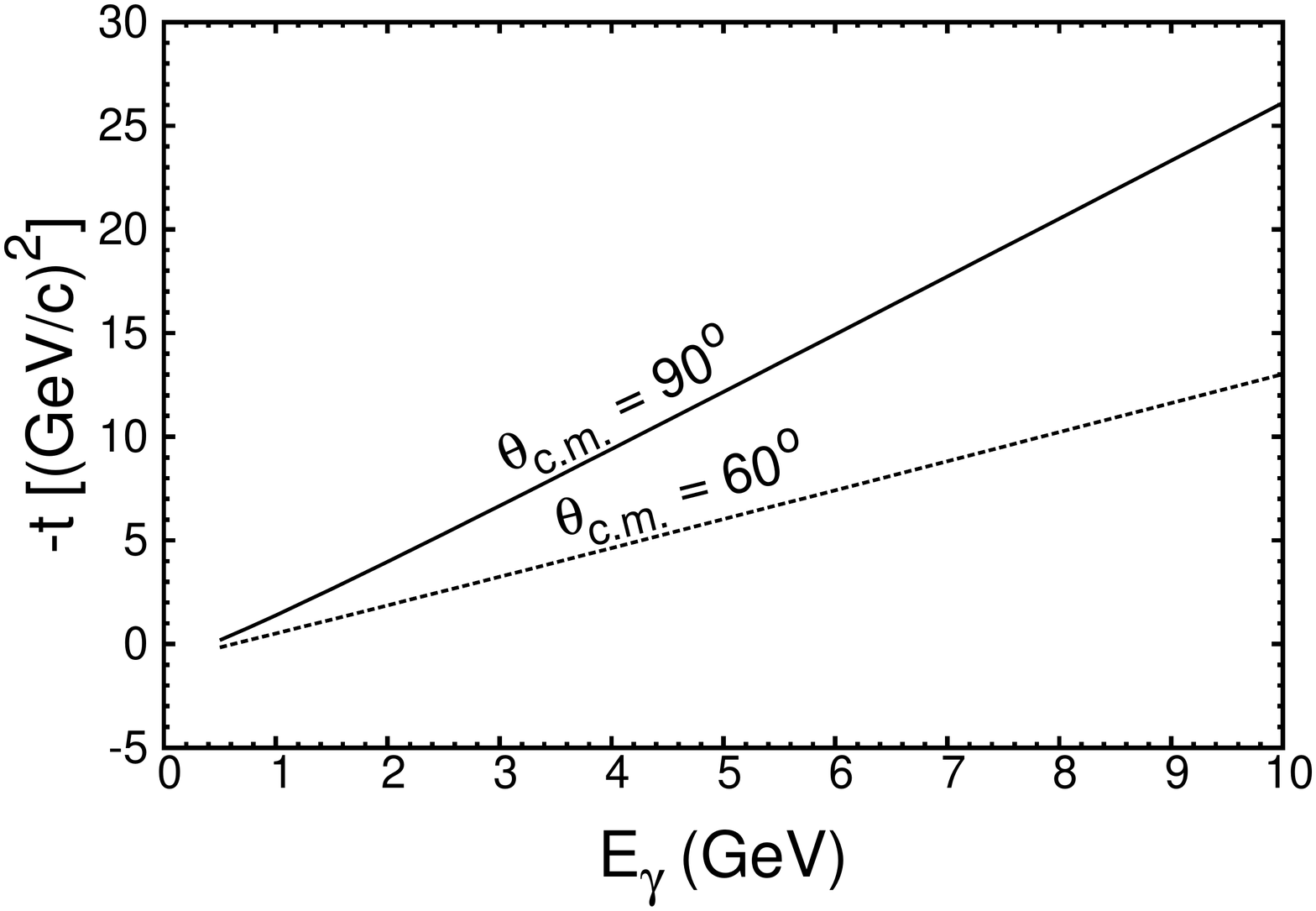}%
}\hfill
\subfloat[\label{FigSub2}]{%
\includegraphics[height = 6 cm, width = 8 cm]{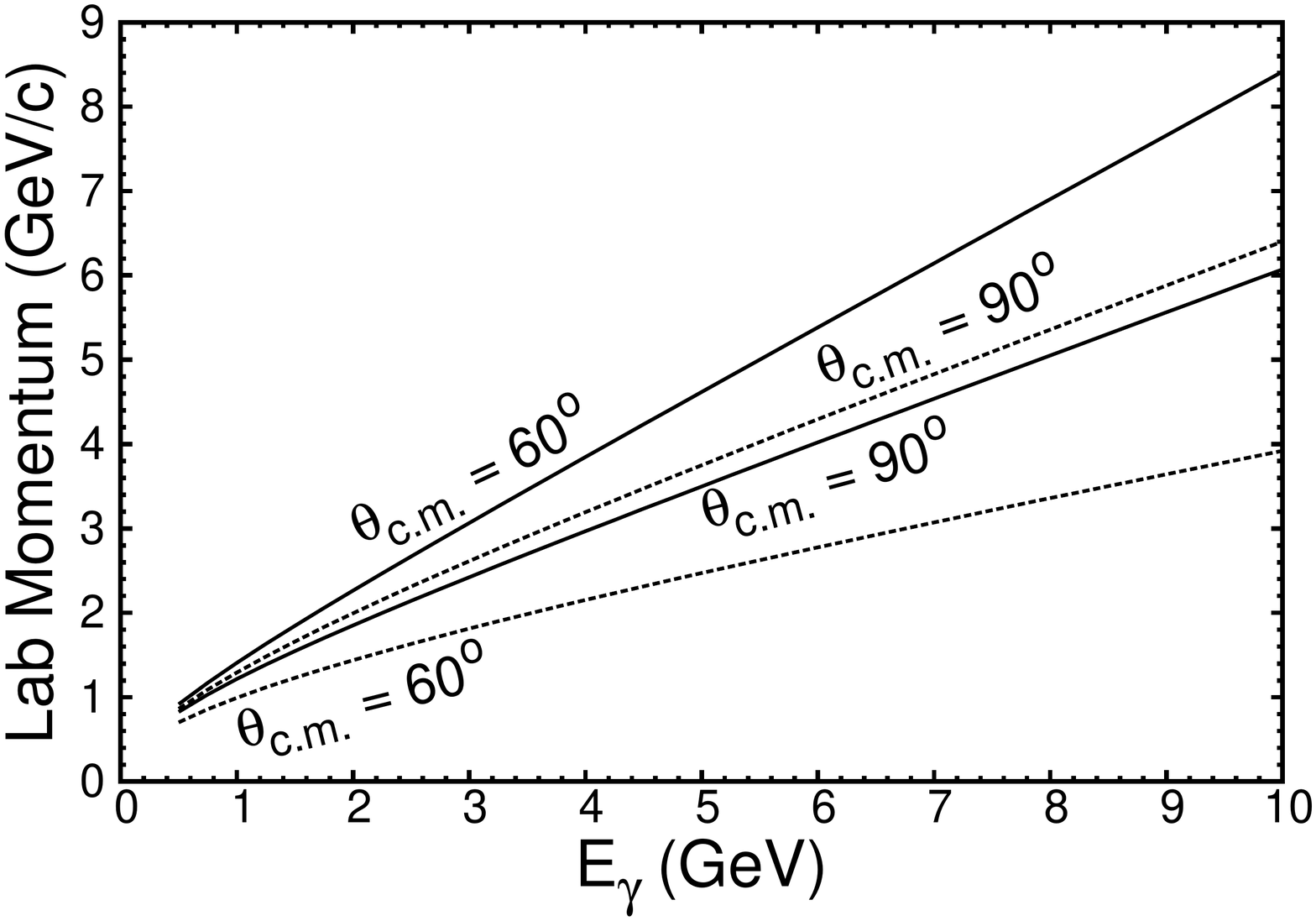}%
}
\captionsetup{format = plain, justification = raggedright }
\caption{ (a) Photon energy dependence of invariant momentum transfer  $-t$. (b)Laboratory momenta of outgoing proton and deuteron as a function of photon energy. Solid lines denote proton, dashed lines denote deuteron.  Calculations are done for 
$\theta_{cm}=90^o$ and $60^o$.}
\label{Figure1}
\end{figure}

That the reaction (\ref{reaction})  at $E_\gamma\gtrsim 1$~GeV and $\theta_{cm}\sim 90^0$ cannot be considered 
a conventional nuclear process with knocked-out nucleon and recoiled residual nuclear system follows
from Fig.1(b), where the  laboratory momenta of outgoing proton and deuteron are given for large $\theta_{cm}$.  
In this case, one  observes that starting at $E_\gamma > 1$~GeV/$c$ the momenta of outgoing proton and  deuteron $> 1$GeV/$c$. 
Such a large momentum of the deuteron significantly exceeds the 
characteristic Fermi momentum  in the $^3$He nucleus, thus the deuteron cannot be considered  residual.  The  momenta of the deuteron are also out of the kinematic range of eikonal, small-angle rescattering\cite{gea,ms01,noredepn}, further diminishing the possibility of describing reaction~(\ref{reaction})  within the framework of conventional nuclear scattering.

Finally, another important feature of the large center-of-mass breakup kinematics is the early onset of QCD degrees of 
freedom due to the large inelasticities (or large masses) produced in the intermediate state of the reaction.  As  shown 
in Ref.\cite{GilmanGross}  for   photodisinegration of the deuteron, already at photon energies of 1~GeV   
one needs  around 15 channels of resonances in the intermediate state 
to describe the process within the hadronic approach.   This situation is similar in the case of the  $^3$He target, in 
which one estimates the produced mass of the intermediate state as $m_R \approx \sqrt{s}- M_d$. From 
this relation one observes that already at $E_\gamma=1$~GeV, $m_R \approx 1.8$~GeV, which is close to the deep inelastic 
threshold of $2$ GeV, for which QCD degrees of freedom are more adequate.
  
Overall, the above kinematical discussion gives  justification for the theoretical description based on 
the QCD degrees of freedom to be increasingly valid starting at photon energies of $\sim 1$~GeV.
\medskip

To conclude the section, we define the reference frame in which the reaction~(\ref{reaction}) will be considered. 
It is defined from the condition for the ``+"  and  transverse components of incoming photon, $q^+ = q_\perp =  0$,  with the 
 photon and target nucleus having the following light-cone four-momenta:
\begin{eqnarray}
 q^\mu & = &  (q_+, q_-, q_\perp) = (0,\sqrt{s'_{^3\text{He}}},0) \nonumber \\
 p^\mu_{^3\text{He}} & = & (p_{^3\text{He}+}, p_{^3\text{He}-},p_{^3\text{He}\perp}) = (\sqrt{s'_{^3\text{He}}}, \frac{m_{^3\text{He}}^2}{\sqrt{s'_{^3\text{He}}}},0),
 \label{InitialMomentaofPhotonHelium}
\end{eqnarray}
where $s'_{^3\text{He}} = s - m_{^3\text{He}}^2$. In the above expression the $\pm$ components are defined as $p_{\pm} = E\pm p_z$, where 
the direction of $z$ axis is opposite to the momentum of the incoming photon in the laboratory frame.

\section{Hard Rescattering Mechanism}
\begin{figure}[ht]
\centering
\includegraphics[width=1.0\textwidth]{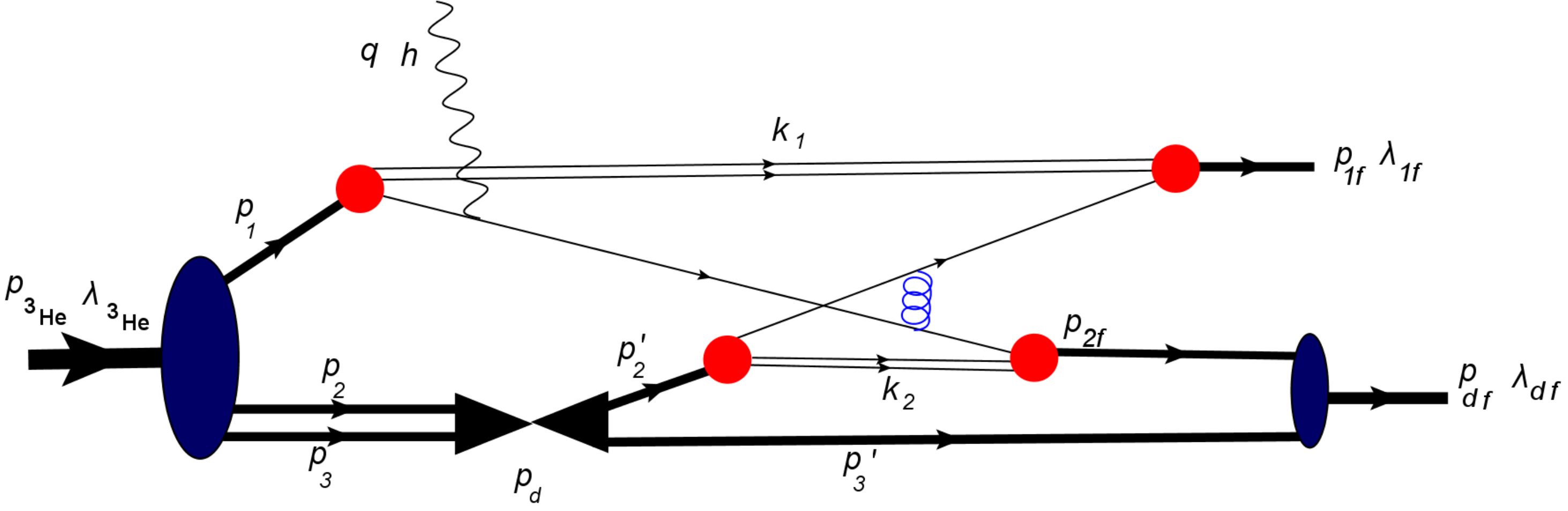}
\caption{Typical diagram of hard rescattering 
mechanism of the $\gamma ^3\text{He}\rightarrow pd$ reaction.}
\label{diagram}
\end{figure}

In the HRM model, the hard photodisintegration takes place in two stages. First, the incoming photon knocks out  a quark from 
one of the nucleons. 
Then in the second step the outgoing fast quark undergoes a high momentum transfer hard scattering with the quark of 
the other nucleon sharing its large 
momentum among the constituents in the final state of the reaction. Since HRM utilizes the small momentum part of the target wave function, which 
has a large component of the initial $pd$ state,  
it is assumed that the energetic photon is  absorbed by any of the quarks belonging to  the protons in the nucleus,  with the subsequent hard rescattering of struck quarks off the quarks in the ``initial" $d$ system  producing the final $pd$  state.
Within such a scenario, the total scattering amplitude can be expressed  as a sum of the multitude of the diagrams similar to that of Fig. \ref{diagram}, with all possibilities of  struck and rescattered quarks combining into a  fast outgoing $pd$ system. Instead of summing  all  the possible diagrams,  
the idea of HRM  is to factorize the hard $\gamma q$ scattering and sum the remaining parts to the amplitude of hard elastic $pd\rightarrow pd$ scattering.  
In this way the complexities related to the large number of diagrams and  nonperturbative quark wave function of the nucleons are absorbed 
into the $pd\rightarrow pd$ amplitude, which  can be taken from experiment.
To demonstrate the above described concept of HRM, we consider the typical 
scattering diagram of Fig. \ref{diagram}. Here, the incoming photon  knocks out a quark from one of the protons in the nucleus. 
The struck quark that now carries almost the whole momentum of the photon will share its momentum with a quark from the other nucleons through 
the quark interchange. The resulting two energetic quarks will recombine with the residual quark-gluon systems to produce a proton and deuteron with 
large relative momentum. 
Note that the  assumption that the nuclear spectator system is represented by intermediate deuteron state is justified based on our previous studies of HRM\cite{gdpn,gheppn}, in which it was found that the scattering amplitude is dominated by small initial momenta of interacting  nucleons. For the case of reaction Eq. (\ref{reaction}), because of the presence of the deuteron in the final state, the small momentum of the initial proton   in the $^3\text{He}$ nucleus will originate predominantly from a two-body $pd$ state. 

In Fig. \ref{diagram}, $h$, $\lambda_{^3\text{He}}, \lambda_{1f} \text{ and } \lambda_{df}$ are the helicities of the incoming photon, $^3\text{He}$ nucleus, and outgoing proton and deuteron respectively. Similarly, $q$, $p_{^3\text{He}}, p_1,p_{1f},p_{d} \text{ and } p_{df}$ are the momenta of the photon, $^3\text{He}$ nucleus, initial and outgoing protons, intermediate deuteron, and the final deuteron respectively. The $k$'s define the momenta of the \textit{spectator} quark systems. The four-momenta defined in Fig. \ref{diagram} satisfy the following relations:
\begin{displaymath}
p_{^3\text{He}} = p_1 + p_2 + p_3;\quad p_2 + p_3 = p_{d} = p_2^{\prime} + p_3^{\prime};\quad 
 p_{2f} + p_{3}^{\prime} = p_{df} ;\quad 
p_{^3\text{He}} + q = p_{1f} + p_{df},
\end{displaymath}
where $p_2$, $p_3$, $p_2^\prime$ and $p_3^\prime$ are four-momenta of the nucleons in the intermediate state deuteron.

We now write the Feynman amplitude 
corresponding to the diagram of Fig. \ref{diagram}, identifying terms corresponding to 
nuclear and nucleonic parts as follows:
\begin{eqnarray} \label{StartDerivationAmplitude}
&&\mathcal{M}^{\lambda_{df},\lambda_{1f};\lambda_{^3\text{He}},h} =  \sum_{\lambda_d'} \int \chi_{d}^{*\lambda_{d}'} (-i\Gamma_{_{DNN}}^{\dagger}) \frac{i(\slashed{p}_{2f} + m)}{p_{2f}^{2} - m_N^2 + i\epsilon} \frac{i(\slashed{p}_{3}'+ m)}{p_{3}'^{2}-m_N^2+i\epsilon} \frac{i(\slashed{p}_{2}' + m)}{p_{2}'^{2}-m_N^2 + i\epsilon} \nonumber \\
&&A:\qquad  i \frac{\Gamma_{_{DNN}} \chi_{d}^{\lambda_{d}} \chi_{d}^{*\lambda_{d}}}{p_{d}^2 - m_{_d}^2 + i\epsilon} (-i)\Gamma_{_{DNN}}^{\dagger} \frac{i(\slashed{p}_{3}+m)}{p_{3}^{2}-m_N^2+i\epsilon} \frac{i(\slashed{p}_{2}+m)}{p_{2}^{2}-m_N^2+i\epsilon} \frac{i(\slashed{p}_{1}+m)}{p_{1}^{2}-m_N^2+i\epsilon}\nonumber \\
&& \qquad i\Gamma_{_{^3\text{He}}} \chi_{_{^3\text{He}}}^{\lambda_{^3\text{He}}} \dfrac{d^{4}p_{2}'}{(2\pi)^{4}} \dfrac{d^{4}p_3}{(2\pi)^{4}} \dfrac{d^{4}p_{3}'}{(2\pi)^{4}} \nonumber \\
&& N1:  \int \chi_{p_{1f}} (-i)\Gamma_{N1}^{\dagger} \frac{i(\slashed{p}_{1f} - \slashed{k}_1 + m)}{(p_{1f} - k_1)^{2} - m_q^{2} + i\epsilon} \Big[-igT_{c}^{\beta} \gamma_{\mu}\Big] \frac{iS(k_{1})}{k_{1}^{2} - m_s^{2} + i\epsilon} \nonumber \\
&& \qquad \frac{i(\slashed{p}_{1} - \slashed{k}_1 + m_q)}{(p_{1} - k_1)^{2} - m_q^{2} + i\epsilon} i\Gamma_{n1}\dfrac{d^{4}k_{1}}{(2\pi)^{4}} \nonumber  \\
&& N2: \int (-i)\Gamma_{N}^{\dagger} \frac{i(\slashed{p}_{2f} - \slashed{k}_2 + m_q)}{(p_{2f} - k_2)^{2} - m_q^{2} + i\epsilon} \frac{iS(k_{2})}{k_{2}^{2} - m_s^{2} + i\epsilon} \frac{i(\slashed{p}_{2}' - \slashed{k}_2 + m_q)}{(p_{2}' - k_2)^{2} - m_q^{2} + i\epsilon}  i\Gamma_{n2'}\dfrac{d^{4}k_{2}}{(2\pi)^{4}} \nonumber \\
&& \gamma: -igT_c^\alpha \gamma_\nu \frac{i(\slashed{p}_1 + \slashed{q} - \slashed{k_1}+m_q)}{(p_1 - k_1 + q)^2 -m_q^2 + i\epsilon} \Big[-ie\gamma^\mu \epsilon^\mu_h\Big] \nonumber \\
&& g: \frac{i d_{\mu \nu}\delta_{\alpha \beta}}{q_{q}^{2}}.
\end{eqnarray}

Here the label  $A$ identifies the nuclear part of the scattering amplitude characterized 
by the transition vertices $\Gamma_{^3\text{He}}$ (for  the  $^3\text{He}\rightarrow N_1,N_2,N_3$ transition) and 
$\Gamma_{DNN}$ (for $D\rightarrow N_2N_3$ transitions). 
The parts $N1$ and $N2$ identify the transition of nucleons $N_1$ and $N_2$ to the quark-spectator system (characterized by the vertex $\Gamma_{N}$) 
with recombination to the final $N_{1f}$ and $N_{2f}$ nucleons.
Here $S(k_1)$ and $S(k_2)$ denote the propagators of the spectator quark-gluons system.  
The label $\gamma$ identifies  the part in which  the photon with polarization  $\epsilon^\mu_h$   
interacts  with  the ($p_1-k_1$) four-momentum   quark  followed by the struck quark propagation.   
The label $g$  represents the gluon propagator. Everywhere, $\chi$'s denote the spin wave functions of the nuclei and nucleons, with $\lambda$'s defining the helicities. The summation over  $\lambda_{d}'$ represents the sum over the helicities of the intermediate deuteron.
The factor $g$ is the QCD coupling  constant with $T_c$ being color matrices.

The hard rescattering model,  which allows us to calculate the sum of the all diagrams similar to Fig. \ref{diagram} is based on the three following assumptions:
\begin{enumerate}
\item The dominant contribution comes from the soft $^3\text{He} \rightarrow pd$ transition defined by small initial momentum of the proton. As a result, this transition can be calculated using nonrelativistic wave functions of the $^3\text{He}$ and deuteron.
\item The high energy $\gamma q$ scattering can be factorized from the final state quark interchange rescattering.
\item  All quark-interchange rescatterings  can be summed into the elastic $pd\rightarrow pd$ amplitude. 
\end{enumerate}

We proceed with the calculation of the amplitude of Eq. (\ref{StartDerivationAmplitude}) by introducing light-cone momenta  $p^\mu = (p_+, p_-, p_\perp)$ 
and also using differentials $d^4p = \frac{1}{2}dp_+dp_-d^2p_\perp$. Furthermore, we perform integrations over the minus component of the momenta.
First, we integrate by $dp_{d-}'$, $dp_{3-}$, and  $d p_{3-}'$  through their pole values in the propagators of the  intermediate deuteron, nucleon $3$, and nucleon $3^\prime$. This allows us to introduce the $pd$ component wave
function of the $^3\text{He}$ [Eq. (\ref{He3wf})]  as well as the $pn$ component deuteron wave function in the intermediate  and final states [Eq. (\ref{DeuteronWF})] of the reaction.
 
In the next step the $dk_{1-}$ and $dk_{2-}$ integrations are performed. 
The $dk_{1-}$  integration allows us to introduce the quark wave functions for nucleons $1$ and $1f$, while the 
$dk_{2-}$ integration does the same  for nucleons $2$ and $2f$. The light-front quark wave function of the nucleon is defined according to Eq. (\ref{QuarkWF}).

After the  ``minus" component integrations and introduction of nuclear and nucleon wave functions, Eq. (\ref{StartDerivationAmplitude})  reduces to

\begin{eqnarray} \label{ReducedAmp}
&&\mathcal{M}^{\lambda_{df},\lambda_{1f};\lambda_{^3\text{He}},h}= \sum_{\substack{  (\lambda_{2f})(\lambda_{2}', \lambda_{3}')(\lambda_{d}) \\ 
(\lambda_1, \lambda_2,\lambda_3 )\\ (\eta_1,   \eta_{2}')(\eta_{1f},\eta_{2f})}}  \int  \frac{\Psi_{d}^{\dagger \lambda_{df}:\lambda_{3}', \lambda_{2f}}
(\alpha_{2f}/\gamma_{d},p_{2\perp},\alpha_{3}'/\gamma_{d},p_{3\perp}')}{1-\alpha_{3}'/\gamma_{d}} 
\Bigg\{ \frac{\Psi_{n2f}^{\dagger \lambda_{2f};\eta_{2f}}(x_{s2},p_{2f\perp},k_{2\perp})}{1-x_{s2}}  \nonumber \\
&&\times \bar{u}_q(p_{2f} - k_2, \eta_{2f}) [-igT_c^\alpha \gamma_\nu]  
\Big[\frac{i(\slashed{p}_1 + \slashed{q} - \slashed{k_1}+m_q)}{(p_1 - k_1 + q)^2 -m_q^2 + i\epsilon}
 \Big] [-ie\epsilon^\mu \gamma\mu] u_q(p_1 - k_1, \eta_1)  \nonumber \\
&& \times \frac{\Psi_{n1}^{\lambda_1;\eta_1}(x_1, k_{1\perp},p_{1\perp}) }{1 - x_1} \Bigg\}_{1} 
 \Bigg\{ \frac{\Psi_{n1f}^{\dagger \lambda_{1f};\eta_{1f}} (x_{s1},k_{1\perp}, p_{1f\perp})}{1 - x_{s1}} 
 \bar{u}_q(p_{1f}-k_1, \eta_{1f}) [-igT_c^\beta\gamma_\mu] u_q(p_{2}'-k_2,\eta_{2}')  \nonumber \\
 && \times
 \frac{\Psi_{n2'}^{\lambda_{2}';\eta_{2}'}(x_{2}',p_{2\perp}',k_{2\perp})}{1-x_{2}'} \Bigg\}_{2} 
 G^{\mu\nu}(r)  \frac{\Psi_d^{\lambda_{d}:\lambda_{2}', \lambda_{3}'}(\alpha_{3}',p_{d\perp},p_{3\perp}')}{1-\alpha_{3}'} 
 \frac{\Psi_{_{d}}^{\dagger \lambda_{d}:\lambda_{2}, \lambda_{3}}(\alpha_{3},p_{3\perp},p_{{d\perp}})}{1-\alpha_{3}}  \nonumber \\
&& \times \frac{\Psi_{_{^3\text{He}}}^{^{\lambda_{^3\text{He}}}}(\beta_1,\lambda_1,p_{_{1\perp}},\beta_2,p_{_{2\perp}}\lambda_2,\lambda_3)}{\beta_1} 
 \frac{d \beta_{d}}{\beta_d} \frac{d^2p_{d\perp}}{2(2\pi)^3} \frac{d \beta_{3}}{\beta_3} \frac{d^2p_{3\perp}}{2(2\pi)^3}
  \frac{d \alpha_{3}'}{\alpha_{3}'} \frac{d^2p_{3\perp}'}{2(2\pi)^3}  \frac{dx_1}{x_1}\frac{d^2k_{1\perp}}{2(2\pi)^3} \frac{dx_{2}'}{x_{2}'}\frac{d^2k_{2\perp}}{2(2\pi)^3},
\end{eqnarray}
where $\beta_i =\frac {p_{i+}} {p_{A+}}$, with $\beta_d, \beta_1,\beta_2$ and $\beta_3$ representing the fractions of the initial light-cone momentum of the $^3\text{He}$ 
nucleus carried by the deuteron and nucleons 1, 2, and 3 respectively. Similarly, $\alpha_i =\frac{ {p_{i+}}}{ p_{d+}}$, with $\alpha_{3}$ and $\alpha_2$ representing 
the momentum fractions of the \textit{intermediate} deuteron carried by the nucleons 3 and 2. The quantity $\gamma_{d}= \frac{p_{df+}}{ p_{d+}}$ is the momentum fraction of the \textit{intermediate} deuteron carried  by the final deuteron. The quantities $x_{1}$ and $x_2$ represent the momentum fractions of the initial nucleons 1 and 2 
carried by the spectator quark system in the corresponding nucleon. The $x_{s1(s2)}$ are the same for the final nucleons 1 (2). The quantities $p_{n\perp}$, 
$p_{n\perp}'$, and $p_{nf\perp}$ with $n = 1,2,3,d$ represent the transverse momenta of nucleons and the deuteron in the initial, intermediate 
and final states of the scattering. The quantities  $k_{1\perp}$ and $k_{2\perp}$
represent the transverse momenta of the spectator quark system in  nucleons 1 and 2 respectively.
The scattering process in Eq. (\ref{ReducedAmp}) can be described in the following blocks:
\begin{itemize}
\item In the initial state, the $^3\text{He}$ wave function describes the transition of the 
$^3\text{He}$ nucleus with helicity $\lambda_{^3\text{He}}$ to the three-nucleon 
intermediate state  with helicities $\lambda_1, \lambda_2, \text{ and } \lambda_3$. The nucleons ``2" and ``3" combine to form an intermediate deuteron, 
which is described by the deuteron wave function.
\item The terms in $\{ ... \}_1$ describe the knocking out of a quark with helicity $\eta_1$ from the proton ``1" by the  photon, with helicity $h$. 
The struck quark then interchanges with a quark from one of the nucleons in the intermediate deuteron state  recombining into the nucleon with helicity 
$\lambda_{2f}$. This nucleon then combines with the nucleon with helicity $\lambda_3$ and produces the final $\lambda_{df}$ helicity deuteron.
\item The terms in $\{ ...\}_2$ describe the emergence of a quark with helicity $\eta_{2}'$  from the $\lambda_{2}'$-helicity nucleon, 
which then interacts with the knocked out quark by exchanging a gluon and  producing a quark with helicity $\eta_{1f}$. This quark then combines 
with the spectator quarks and produces a final nucleon with helicity $\lambda_{1f}$.
\end{itemize}
To proceed with the calculation of the amplitude in Eq. (\ref{ReducedAmp}), we first  identify the pole in the denominator of the propagator of the knock-out quark, as follows:
\begin{eqnarray} \label{PoleValueofPropagator}
&&(p_1 - k_1 + q )^2 - m_q^2 + i\epsilon = s'_{^3\text{He}}(1-x_1)(\beta_1 - \beta_s + i\epsilon), \nonumber \\
&&\text{ where } \beta_s =  -\frac{1}{s'_{^3\text{He}}} \Big( m_N^2 + p_{1\perp}^{2} - \frac{m_{s}^{2} + k_{1\perp}^{2}}{x_1} - 
\frac{m_q^2 + (p_{1\perp} - k_{1\perp})^2 }{1-x_1} \Big).
\end{eqnarray}
From this point onward, our discussion is based on the fact that the $^3\text{He}$ wave function strongly peaks at $\beta_1 = \beta_s = \frac{1}{3}$. 
This corresponds to the kinematic situation in which the nucleons in  $^3\text{He}$ have small momentum and as a result they share equal 
amounts of momentum fractions of the nucleus.  In the following calculations we will estimate the integral in Eq. (\ref{ReducedAmp}) 
at the pole value of the propagator (\ref{PoleValueofPropagator}). 
This justifies the use of the sum rule $\sum \limits_{\lambda} u(p, \lambda)\bar u(p, \lambda) = \slashed{p} +m$ for the numerator of the struck quark propagator, resulting in
\begin{eqnarray} \label{AmplitudeAfterPoleValuePropagatorIntroduced}
&&\mathcal{M}^{\lambda_{df},\lambda_{1f};\lambda_{^3\text{He}},h}=  \sum_{\substack{ (\lambda_{2f})(\lambda_{2}' ,\lambda_{3}') 
(\lambda_{d})\\ (\lambda_1, \lambda_2,\lambda_3) \\ (\eta_1, \eta_{q1}) (\eta_{1f}, \eta_{2f})(\eta_{2}')}}  \int  
\frac{{\Psi_{d}^{\dagger \lambda_{df}:\lambda_{3}', \lambda_{2f}}}(\alpha_{2f}/\gamma_{d},p_{2\perp},\alpha_{3}'/\gamma_{d},p_{3\perp}')}{1-\alpha_{3}'/\gamma_{d}}  \nonumber \\
&& \times \Bigg\{ \frac{\Psi_{n2f}^{\dagger \lambda_{2f};\eta_{2f}}(x_{s2},p_{2f\perp},k_{2\perp})}{1-x_{s2}} 
\bar{u}_q(p_{2f}-k_2, \eta_{2f})
 [-igT_c^\alpha \gamma_\nu] 
 \Big[\frac{u_q(p_1+q-k_1,\eta_{q1}) \bar{u}_q(p_1+q-k_1,\eta_{q1}) }{s'(1-x_1)(\beta_1 - \beta_s + i\epsilon)} \Big]   \nonumber \\
 && \times [-ie\epsilon^\mu \gamma_\mu]  u_q(p_1 - k_1, \eta_1) 
\frac{\Psi_{n1}^{\lambda_1;\eta_1}(x_1, k_{1\perp},p_{1\perp}) }{1 - x_1} \Bigg\}_{1} 
\Bigg\{ \frac{\Psi_{n1f}^{\dagger \lambda_{1f};\eta_{1f}} (x_{s1},k_{1\perp}, p_{1f\perp})}{1 - x_{s1}} \bar{u}_q(p_{1f}-k_1, \eta_{1f})  \nonumber \\
&& \times [-igT_c^\beta\gamma_\mu] u_q(p_{2}'-k_2,\eta_{2}')
\frac{\Psi_{n2'}^{\lambda_{2}';\eta_{2}'}(x_{2}',p_{2\perp}',k_{2\perp})}{1-x_{2}'} \Bigg\}_{2} G^{\mu\nu}(r) \frac{\Psi_d^{\lambda_{d}:\lambda_{2}', \lambda_{3}'}(\alpha_{3}',p_{d\perp},p_{3\perp}')}{1-\alpha_{3}'} \nonumber \\
&& \times \frac{\Psi_{_{d}}^{\dagger \lambda_{d}:\lambda_{2}, \lambda_{3}}(\alpha_{3},p_{_{3\perp}},p_{_{d\perp}})}{1-\alpha_{3}}  
\frac{\Psi_{_{^3\text{He}}}^{^{\lambda_{^3\text{He}}}}(\beta_1,\lambda_1,p_{_{1\perp}},\beta_2,p_{_{2\perp}}
\lambda_2,\lambda_3)}{\beta_1}  \frac{d \beta_{d}}{\beta_d} \frac{d^2p_{d\perp}}{2(2\pi)^3}\frac{d \beta_{3}}{\beta_3} \frac{d^2p_{3\perp}}{2(2\pi)^3}  \nonumber \\
&& \times \frac{d \alpha_{3}'}{\alpha_{3}'} \frac{d^2p_{3\perp}'}{2(2\pi)^3}
 \frac{dx_1}{x_1}\frac{d^2k_{1\perp}}{2(2\pi)^3} \frac{dx_{2}'}{x_{2}'}\frac{d^2k_{2\perp}}{2(2\pi)^3}.
\end{eqnarray}
In Eq. (\ref{AmplitudeAfterPoleValuePropagatorIntroduced}), using the relations $\beta_1 + \beta_d = 1$ and 
$d\beta_d = d\beta_1$, we perform integration by 
$d\beta_1$ estimating it at the pole,  $\beta_1 = \beta_s$.   For this we express
 \begin{equation} \label{IntegralWithPVpart}
 \frac{1}{\beta_1 -{\beta_s} + i\epsilon} = -i\pi \delta(\beta_1-{\beta_s}) + \text{P.V.} \int \frac{d\beta_1}{\beta_1 - {\beta_s}},
\end{equation}
and neglect the principal value (P.V.)  part since its contribution is defined by the nuclear wave function at internal momenta of $\sim \sqrt{s}$  
and is strongly suppressed (see, e.g., Refs. [8,9]). Restricting by the first term of Eq. (\ref{IntegralWithPVpart}) 
allows us to use the on-shell approximation to calculate the matrix element of the photon-quark interaction. 
Using the relation, $(p_1 -k_1)_+\gg k_\perp, m_q$ for the matrix element, one obtains (for details see Appendix A)
\begin{equation} \label{PhotonQuarkInteractionMatrixElement}
\bar{u}_q (p_1 - k_1 + q,\eta_{q1}) [ie\epsilon^\perp \gamma^\perp]u_q(p_1 - k_1, \eta_{1}) = 
ie Q_i2 \sqrt{2E_1E_2}(-h) \delta^{\eta_{q1} h} \delta^{\eta_1 h},
\end{equation}
where $E_1 =\frac{\sqrt{s'_{^3\text{He}}}}{2} \beta_1 (1 - x_1)$ \text{ and } $ E_2 = \frac{\sqrt{s'_{^3\text{He}}}}{2}[1 - \beta_1(1 - x_1)]$ 
are the energies of the  struck quark before and after the interaction with the photon. 
The factor $Q_{i}$ is the charge of the struck quark in $e$ units. 
The above result indicates that the incoming  $h$-helicity  photon selects the quark with the same helicity ($h=\eta_1$), conserving it during the interaction ($h = \eta_{q_1}$). 
The above integration sets $\beta_1=  \beta_s$ and $\beta_d = 1-  \beta_s$. To proceed, using the fact that  the $^3\text{He}$ wave function peaks at $\beta_s = \frac{1}{3}$, we apply the ``peaking" approximation 
in which the integrand of Eq. (\ref{AmplitudeAfterPoleValuePropagatorIntroduced}) is estimated  at  $\beta_1 = \beta_s = \frac{1}{3}$ 
and $\beta_d = \frac{2}{3}$. Moreover, it follows from Eq. (\ref{PoleValueofPropagator}), the 
$\beta_s = {1/3}$ condition restricts $x_1\sim  \frac{m_s^2}{ s}$.  The latter condition allows us to simplify further the matrix element in 
Eq. (\ref{PhotonQuarkInteractionMatrixElement}) approximating $E_1\approx \frac{\sqrt{s^\prime}} {6}$ and  $E_2 \approx \frac{\sqrt{s^\prime}}{3}$. 
This results in
\begin{eqnarray} \label{AmpAfterPhotonQuarkMatrixElement}
&& \mathcal{M}^{\lambda_{df},\lambda_{1f};\lambda_{^3\text{He}},h}=  \frac{3 }{4} (-h)\frac{1}{\sqrt{s'_{^3\text{He}}}} \sum_{i}eQ_i
 \sum_{\substack{(\lambda_{2f})(\lambda_{2}',\lambda_{3}') (\lambda_{d}) \\ (\lambda_1,\lambda_2,\lambda_3) \\ 
 (\eta_{1f}, \eta_{2f}) (\eta_{2}' ) }} 
 \int  \frac{{\Psi_{d}^{\dagger \lambda_{df}:\lambda_{3}^\prime,\lambda_{2f}}}(\alpha_{2f}
 /\gamma_{d},p_{2\perp},\alpha_{3}'/\gamma_{d},p_{3\perp}')}{1-\alpha_{3}'/\gamma_{d}} \nonumber \\
 && \times \Bigg\{ \frac{\Psi_{n2f}^{\dagger \lambda_{2f};\eta_{2f}}(x_{s2},p_{2f\perp},k_{2\perp})}{1-x_{s2}}  
\bar{u}_q(p_{2f}-k_2, \eta_{2f})
[-igT_c^\alpha \gamma_\nu] 
 \Big[u_q(p_1+q-k_1,h)   \Big]  \nonumber \\
&& \times \frac{\Psi_{n1}^{\lambda_1;h}(x_1, k_{1\perp},p_{1\perp}) }{1 - x_1} \Bigg\}_{1}
 \Bigg\{ \frac{\Psi_{n1f}^{\dagger \lambda_{1f};\eta_{1f}} (x_{s1},k_{1\perp}, p_{1f\perp})}{1 - x_{s1}}
 \bar{u}_q(p_{1f}-k_1, \eta_{1f}) [-igT_c^\beta\gamma_\mu]  \nonumber \\
&& \times u_q(p_{2}'-k_2,\eta_{2}')
 \frac{\Psi_{n2'}^{\lambda_{2}';\eta_{2}'}(x_{2}',p_{2\perp}',k_{2\perp})}{1-x_{2}'} \Bigg\}_{2} G^{\mu\nu}(r) 
 \frac{\Psi_d^{\lambda_{d}:\lambda_{2}',\lambda_{3}'}(\alpha_{3}',p_{d\perp},p_{3\perp}')}{1-\alpha_{3}'} 
 \frac{\Psi_{_{d}}^{\dagger \lambda_{d}:\lambda_{2},\lambda_{3}}(\alpha_{3},p_{_{3\perp}},p_{_{d\perp}})}{1-\alpha_{3}}  \nonumber \\
 &&\times \Psi_{_{^3\text{He}}}^{^{\lambda_{^3\text{He}}}}(\beta_1=1/3,\lambda_1,p_{_{1\perp}},\beta_2,p_{_{2\perp}}\lambda_2,\lambda_3) 
  \frac{d^2p_{d\perp}}{(2\pi)^2} 
\frac{d \beta_{3}}{\beta_3} \frac{d^2p_{3\perp}}{2(2\pi)^3}\frac{d \alpha_{3}'}{\alpha_{3}'} \frac{d^2p_{3\perp}'}{2(2\pi)^3} \frac{dx_1}{x_1}\frac{d^2k_{1\perp}}{2(2\pi)^3} \frac{dx_{2}'}{x_{2}'}\frac{d^2k_{2\perp}}{2(2\pi)^3}. \nonumber \\
\end{eqnarray} 
The above expression corresponds to the amplitude of Fig. \ref{diagram}.  To be able to calculate the total amplitude 
of $\gamma ^3\text{He} \rightarrow pd$ scattering, one needs to sum the multitude of similar diagrams representing all possible 
combinations of photon coupling to quarks in one of the protons, followed by quark interchanges or possible multi gluon exchanges between outgoing nucleons, producing the final $pd$ system with large relative momentum. The latter rescattering is inherently nonperturbative. The same is true for the quark wave function of the nucleon, which is largely unknown.  
The main idea of HRM is that, instead of calculating all the amplitudes explicitly,  we notice that the hard kernel in 
Eq. (\ref{AmpAfterPhotonQuarkMatrixElement}), 
$ \{\cdots \}_1 \{\cdots\}_2$,  together with the gluon propagator is similar to that of the  hard $pd \rightarrow pd$ scattering. To illustrate this, in Appendix B we calculated the amplitude of hard $pd\rightarrow pd$ scattering corresponding to the diagram of Fig.\ref{Fig_B1}. Using the notations similar to ones used in Fig. \ref{diagram} and the derivation analogous to the above derivation in which  light-front  wave functions of  the deuteron and nucleons are introduced, one arrives at Eq. (\ref{pd_amplitude}).

Equation (\ref{pd_amplitude}) is derived  in the $pd$ center of mass reference frame, in which the final momenta $p_{1f}$ and $p_{df}$ are 
chosen to be  the same as in reaction (\ref{reaction}).
Thus the  $pd\rightarrow pd$ amplitude  is defined at the same $s = (p_{1f} + p_{df})^2$ as  in Eq. (\ref{mandelstams})  
but at the different invariant momentum transfer defined as: $t_{pd} = (p_{df} - p_{d})^2$. 

To be able to substitute Eq. (\ref{pd_amplitude}) into Eq. (\ref{AmpAfterPhotonQuarkMatrixElement}), we notice that within the peaking approximation  the momentum transfer $t_N$ entering in the rescattering part of the amplitude in Eq. (\ref{AmpAfterPhotonQuarkMatrixElement}) is approximately equal to $t_{pd}$:
\begin{equation} 
t_N  \approx t_{pd}  =   (p_{df} - p_d)^2,
\label{t_N}
\end{equation}
where $p_d$ is the deuteron four-momentum in the intermediate state of the reaction (Fig.\ref{diagram}).

Furthermore, due to $q_+=0$, the spinor $u_q(p_1-k_1 + q,h)$ in Eq. (\ref{AmpAfterPhotonQuarkMatrixElement}) is defined at the same momentum fraction $1-x_1$  and transverse momentum as the spinor $u_q(p_1 - k_1)$ in Eq. (\ref{pd_amplitude}). 
The final step that allows us to replace the quark-interchange part of  Eq. (\ref{AmpAfterPhotonQuarkMatrixElement}) by  the 
$pd\rightarrow pd$ amplitude is the observation that due to  the condition of  $\beta_1 = \beta_s \approx \frac{1}{3}$,  it follows from 
Eq. (\ref{PoleValueofPropagator}) that the momentum fraction of the struck quark $1-x_1 \sim 1 - \frac{m_s^2}{ s^\prime_{^3\text{He}}} \sim 1$. 
This justifies the additional assumption according to which 
the helicity of  the struck quark is the same as the nucleon's from which it originates, i.e. $\eta_1 = \lambda_1$.  
With this assumption one can sum 
over $\eta_1$ in Eq. (\ref{pd_amplitude}), which allows us now to substitute it into Eq. (\ref{AmpAfterPhotonQuarkMatrixElement}), yielding
\begin{eqnarray} \label{simplifiedAmplitude}
&\mathcal{M}^{\lambda_{df},\lambda_{1f};\lambda_{^3\text{He}},h} =&  \frac{3}{4}\frac{1}{\sqrt{s'_{^3\text{He}}}} \sum_{i}eQ_i (h)
 \sum_{\substack{\lambda_{d} \\  \lambda_2,\lambda_3 }} \int  \mathcal{M}_{pd}^{\lambda_{df},\lambda_{1f} ;\lambda_{d},h} (s, t_N) \frac{\Psi_{_{d}}^{\dagger \lambda_{d}:\lambda_2,\lambda_3}(\alpha_{3},p_{_{3\perp}},\beta_d,p_{_{d\perp}})}{1-\alpha_{3}}  \nonumber \\
&& \times \qquad \qquad \Psi_{_{^3\text{He}}}^{{\lambda_{^3\text{He}}}: h, \lambda_2, \lambda_3}(\beta_1=1/3,  p_{_{1\perp}},\beta_2,p_{_{2\perp}})
\frac{d^2p_{d\perp}}{(2\pi)^2}\frac{d \beta_3}{\beta_{3}} \frac{d^2p_{3\perp}}{2(2\pi)^3}.
\end{eqnarray}

We can further simplify this  equation using the fact that the  momentum transfer in the $pd\rightarrow pd$ scattering amplitude significantly exceeds the momenta of bound nucleons in the nucleus. As a result, one can factorize the  $pd\rightarrow pd$ amplitude from the integral in Eq. (\ref{simplifiedAmplitude}) at $t_{pd}$ approximated as
\begin{equation}
t_{pd} \approx  [p_{df} - m_d)^2 = (p_{1f} - (m_N+q)]^2,
\label{tpd_fact}
\end{equation}
resulting in
\begin{eqnarray} 
\mathcal{M}^{\lambda_{df},\lambda_{1f};\lambda_{^3\text{He}},h} &=  &  \frac{3}{4} \frac{1}{\sqrt{s'_{^3\text{He}}}} \sum_{i} \sum_{\substack{\lambda_{d} \\  }}eQ_i (h) \mathcal{M}_{pd}^{\lambda_{df},\lambda_{1f} ;\lambda_{d},h} (s, t_{pd})  \nonumber \\
& &  \ \ \ \ \ \times  \int   \Psi_{^3\text{He}/d}^{ \lambda_{^3\text{He}}:\lambda_1,\lambda_d}(\beta_1 = 1/3, p_{1\perp})  \frac{d^2p_{1\perp}}{(2\pi)^2},
\label{pdAmplitudeFactored}
\end{eqnarray}
where we introduced  the light-front nuclear transition wave function as
\begin{eqnarray}
\Psi^{\lambda_{^3\text{He}}:\lambda_1, \lambda_{d}}_{^3\text{He}/d}(\beta_1, p_{1\perp}) &=  & \sum_{\lambda_2, \lambda_3} \int \frac{\Psi_{_{d}}^{ \dagger\lambda_{d}: \lambda_2, \lambda_3}  (\alpha_3, p_{3\perp}, \beta_d, p_{d\perp} )} {2 (1 - \alpha_3)} \Psi_{_{^3\text{He}}}^{{\lambda_{^3\text{He}}}: \lambda_1, \lambda_2, \lambda_3}(\beta_1, p_{1\perp}, \beta_2, p_{2\perp}) \nonumber \\
& & \ \ \ \ \ \ \ \  \times  \frac{d\beta_3}{\beta_3}\frac{d^2p_{3\perp}}{2(2\pi)^3}.	
 \label{LCNTF}
\end{eqnarray}
The above function defines the probability amplitude of the $^3\text{He}$  nucleus 
transitioning to a proton and  deuteron with respective momenta  $p_1$ and $p_d$ and  helicities $\lambda_1$ and $\lambda_d$.

In Eq. (\ref{pdAmplitudeFactored}) one sums over all the valence quarks in the bound proton that interact with incoming photon. 
To calculate such a sum one needs an underlying model for hard nucleon interaction based on the explicit quark degrees of freedom. 
Such a model will allow us to simplify further the amplitude of  Eq. (\ref{pdAmplitudeFactored}), representing it 
through the product of an effective charge $Q_{eff}$ that the incoming photon probes in the reaction and 
the hard $pd \rightarrow pd$ amplitude in the form
\begin{equation} 
\mathcal{M}^{\lambda_{df},\lambda_{1f};\lambda_{^3\text{He}},h} =   \frac{3}{4} \frac{eQ_{eff}(h)}{\sqrt{s'_{^3\text{He}}}}  \sum_{\substack{\lambda_{d} \\  }} \mathcal{M}_{pd}^{\lambda_{df},\lambda_{1f} ;\lambda_{d},h} (s, t_{pd}) 
  \int   \Psi_{^3\text{He}/d}^{ \lambda_{^3\text{He}}:\lambda_1,\lambda_d}
(\beta_1 = 1/3, p_{1\perp}) 
 \frac{d^2p_{1\perp}}{(2\pi)^2}.
\label{Amplitude}
\end{equation}

\section{The Differential Cross Section}

The  differential cross section of  reaction (\ref{reaction}) can be presented in the standard form
\begin{equation} \label{CS}
\frac{d\sigma}{dt} = \frac{1}{16\pi} \frac{1}{s'^2_{^3\text{He}}} |\overline{\mathcal{M}}|^2,
\end{equation}
where, for the case of unpolarized scattering,
\begin{equation} \label{squard_averaged_amp}
|\overline{\mathcal{M}}|^2 = \frac{1}{2} \frac{1}{2} \sum_{\lambda_{^3\text{He}}, h} \sum_{\lambda_{df}, \lambda_{1f}} \left|\mathcal{M}^{\lambda_{df},\lambda_{1f};\lambda_{^3\text{He}},h} \right|^2.
\end{equation}
Here squared amplitude is summed by the final helicities and averaged by the helicities of $^3\text{He}$ and the incoming photon.
The factorization approximation of Eq. (\ref{Amplitude}) allows us to express Eq. (\ref{squard_averaged_amp}) through the
convolution of the averaged square of $pd \rightarrow pd$ amplitude, $\mathcal{M}_{pd}$, in the form

\begin{equation} \label{squaredAmpWithSquaredpdAmp}
|\overline{\mathcal{M}}|^2 = \frac{9}{16} \frac{e^2Q_{eff}^2}{s'_{^3\text{He}}} \frac{1}{2} |\overline{\mathcal{M}_{pd}}|^2 S_{^3\text{He}/d}(\beta_1 = 1/3),
\end{equation}
where
\begin{equation} \label{M_pd2def}
|\overline{\mathcal{M}_{pd}}|^2 = \frac{1}{3} \frac{1}{2} \sum_{\lambda_{df}, \lambda_{1f}; \lambda_d, \lambda_1} \left|\mathcal{M}_{pd}^{\lambda_{df}, \lambda_{1f}; \lambda_d, \lambda_1}(s, t_{pd})\right|^2,
\end{equation}
and the nuclear light-front transition spectral function is defined as
\begin{equation} \label{LCTSF}
S_{^3\text{He}/d} (\beta_1) = \frac{1}{2} \sum_{\lambda_{^3\text{He}}; \lambda_1, \lambda_d} \left|\int \Psi_{^3\text{He}/d}^{\lambda_{^3\text{He}}:\lambda_1, \lambda_d} (\beta_1, p_{1\perp}) \frac{d^2p_{1\perp}}{(2\pi)^2} \right|^2.
\end{equation}
Substituting Eq. (\ref{squaredAmpWithSquaredpdAmp}) into (\ref{CS}), one can express the differential cross section through the
differential cross section of elastic $pd \rightarrow pd$ scattering in the form
\begin{equation} \label{differentialCS}
\frac{d\sigma}{dt} = \frac{9}{32} \frac{e^2Q_{eff}^2}{s'_{^3\text{He}}} \Big(\frac{s'_N}{s'_{^3\text{He}}}\Big) \frac{d\sigma_{pd}}{dt}(s, t_{pd}) S_{^3\text{He}/d}(\beta_1 = 1/3),
\end{equation}
where $s'_N = s - m_N^2$.

\subsection{Numerical estimates of the cross section}

\subsubsection{Calculation of the light-front transition spectral function}
For calculation of the light-front  transition spectral function of Eq. (\ref{LCTSF}) we observe, that 
within the applied peaking approximation which maximizes the nuclear wave function's contribution to the 
scattering amplitude, $\beta_1 = \frac{1}{ 3}$ and $\beta_d \approx \frac{2}{3}$. These values of light-cone momentum 
fractions correspond to a small internal momenta of the nucleons in the nucleus. Additionally since the deuteron wave 
function strongly peaks at small relative momenta between two spectator (``2" and ``3" in Fig. \ref{diagram})  nucleons, 
the  integral in Eq. (\ref{LCNTF}) is dominated at $\beta_3 \approx \frac{1}{3}$ and $\alpha_3\approx \frac{1}{2}$. 
This justifies the application of non-relativistic approximation in the calculation of the transition spectral function of Eq. (\ref{LCTSF}).

In nonrelativistic limit, using the  boost invariance of the momentum fractions $\beta_i$ ($i=1,2,3$),  one relates 
them to the three-momenta  of the constituent nucleons in the laboratory frame of the nucleus as follows:
\begin{equation}
\beta_i  = \frac{p_{i+}}{p_{^3\text{He}+}} \approx \frac{1}{3} + \frac{p^{lab}_{i,z}}{ 3 m_N}.
\label{beta_pz}
\end{equation}

Using above relations one approximates  $\frac{d\beta_3 }{\beta_3}\approx \frac{dp^{lab}_{3z}}{ m_N}$ and 
$\alpha_3\approx \frac{1}{2} + \frac{p_{2,z}}{2 m_N}$ in Eq. (\ref{LCNTF}).
Introducing also  the relative three-momentum in the $2,3$ nucleon system as
 \begin{equation}
\vec{p}_{rel} = \frac{1}{2}(\vec p^{\ lab}_3 - \vec p^{\ lab}_2),
\end{equation}
and using the relation between light-front and non-relativisitc nuclear wave functions in the small-momentum limit (see Appendix C),
\begin{equation}
\Psi^{LC}_A(\beta,p_\perp) = \frac{1}{\sqrt{A}}  \big(m_N 2(2\pi)^3 \big)^\frac{A-1}{ 2}\Psi_{A}^{NR}(\vec{p}),
\end{equation}
one can express the light-cone nuclear transition wave function of Eq. (\ref{LCNTF}) through the 
nonrelativistic  $^3\text{He}$ to $d$ transition wave function as follows:
\begin{equation}
\Psi^{\lambda_{3He}:\lambda_1, \lambda_d}_{^3\text{He}/d}(\beta_1, p_{1\perp})  =
\sqrt{\frac{1}{6}} \sqrt{m_N 2(2\pi)^3} \cdot
\Psi^{\lambda_{3He}:\lambda_1, \lambda_{d}}_{^3\text{He}/d,NR}(\vec p_1),
\label{LCtoNRwft}
\end{equation}
where the nonrelativistic transition wave function is defined as
\begin{equation}
\Psi^{\lambda_{^3\text{He}}:\lambda_1, \lambda_d}_{^3\text{He}/d,NR}(\vec p_1) 
= \sum_{\lambda_2, \lambda_3} \int 
\Psi_{_{d},NR}^{ \dagger\lambda_{d}: \lambda_2, \lambda_3}  (p_{rel}) 
\Psi_{_{^3\text{He}},NR}^{{\lambda_{^3\text{He}}}: \lambda_1, \lambda_2, \lambda_3}(p_{1}, p_{rel}) d^3p_{rel}.	
\label{NRtranswf}
\end{equation}
Using Eq. (\ref{LCtoNRwft}), we  express the light-front spectral function through the nonrelativistic counterpart in the form
\begin{equation}\label{LCspectralNR}
S_{^3\text{He}/d} (\beta_1) = \frac{ m_N 2(2\pi)^3}{6} N_{pd} \ S_{^3\text{He}/d}^{NR} (p^{lab}_{1z}),
\end{equation}
where $\beta_1$ and $p_{1z}^{lab}$ are related according to Eq. (\ref{beta_pz}) 
and $N_{pd}=2$ is the number of the effective $pd$ pairs.  The non-relativistic spectral function is 
defined as
\begin{equation}
S_{^3\text{He}/d}^{NR} (p_{1z}) =  
\frac{1}{2} \sum_{\lambda_{^3\text{He}}; \lambda_1, \lambda_d} \left| 
\int \Psi_{^3\text{He}/d,NR}^{\lambda_{^3\text{He}}:\lambda_1, \lambda_d} (p_{1z}, p_{1\perp}) 
\frac{d^2p_{1\perp}}{(2\pi)^2}\right| ^2,
\end{equation}
where both $^3\text{He}$ and $d$ wave functions are renormalized to unity.
In the above expressions all the momenta  entering in the 
non-relativisitic wave functions are considered in the laboratory frame of the $^3\text{He}$ nucleus.

\subsubsection{Hard elastic $pd\rightarrow pd$ scattering cross section}
The hard  $ pd \rightarrow pd $ elastic scattering cross section entering in Eq. (\ref{differentialCS}) is 
defined at the same invariant energy $s$ as the reaction (\ref{reaction}) but at different [from Eq. (\ref{mandelstams})]
 invariant momentum transfer, $t_{pd}$, defined in Eq. (\ref{tpd_fact}). Comparing Eqs.(\ref{tpd_fact}) and (\ref{mandelstams}), one  can relate the $t_{pd}$ to $t$ in the following form:
\begin{equation} \label{t_pd}
t_{pd} = \frac{1}{3} m_d^2 - \frac{2}{9} m_{^3\text{He}}^2 + \frac{2}{3} t.
\end{equation}
%\begin{equation} \label{cmangles}
%\frac{1}{3} (1 - \cos\theta_{cm}) \approx  \frac{1}{2} (1 - \cos\theta_{cm}^*).
%\end{equation}
\begin{figure}[ht]
\centering
\includegraphics[width = 12cm, height =8cm]{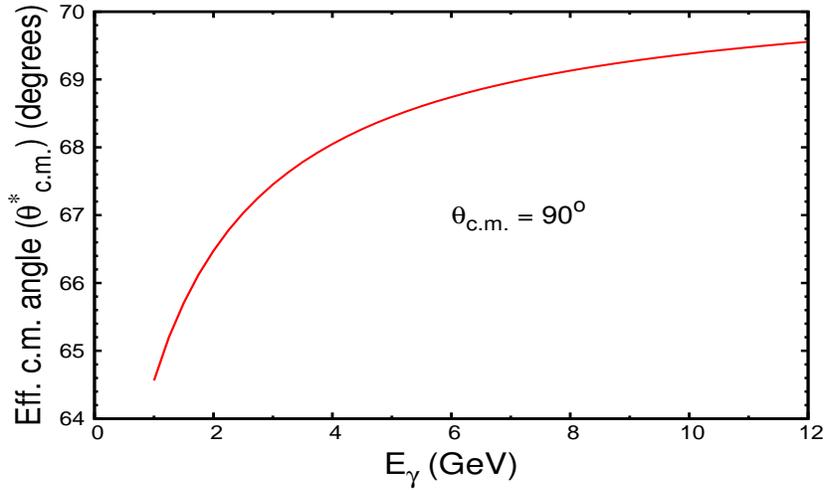}
\caption{Effective center of mass angle vs the incident photon energy.}
\label{eff_cm_angle}
\end{figure} 
It follows from  the above equation, for large momentum transfer,  that, therefore for the same 
$s$, the $pd\rightarrow pd$ 
scattering will take place at smaller angles in the $pd$ center-of-mass reference frame.  To evaluate this difference we 
introduce $\theta_{cm}^*$ which represents the center-of-mass scattering angle for the $pd\rightarrow pd$ reaction in the form
\begin{equation}\label{t_pd_explicit}
t_{pd} = (p_{df} - p_{di})^2 = 2 (m_d^2 - E_{d,cm}^2)(1 - \cos\theta_{cm}^*), 
\end{equation} 
where $E_{d,cm}  = \frac{s+m_{d}^2 - m_N^2}{ 2\sqrt{s}}$. 
Then, comparing this equation with Eq. (\ref{expr_for_t}) in  the asymptotic limit of 
high energies, one finds that for $\theta_{cm} = 90^o$ in reaction (\ref{reaction})  the asymptotic limit of the effective center-of-mass scattering angle of  $pd\rightarrow pd $ scattering is $\theta_{cm}^* = 70.53^o$. 
The dependence  of $\theta_{cm}^*$  at finite energies on the incoming photon is shown in  Fig. \ref{eff_cm_angle}.  The figure indicates that, for realistic comparison of HRM prediction with the data, one needs the $pd\rightarrow pd$ cross section for the range of center-of-mass scattering angles.

\begin{figure} [ht]
\centering
\subfloat[\label{pdfit1}] {%
\includegraphics[height = 7 cm, width = 8 cm]{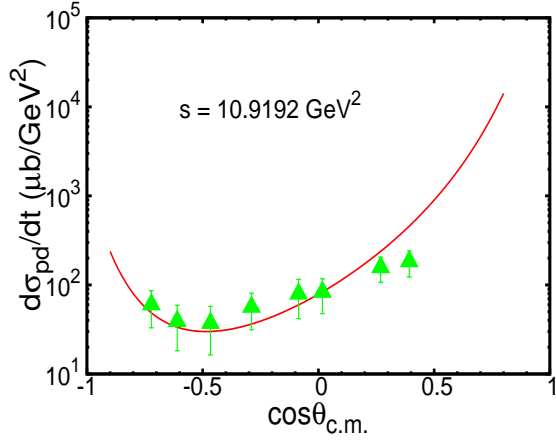}%
}\hfill
\subfloat[\label{pdfit2}] {%
\includegraphics[height = 7 cm, width = 8 cm]{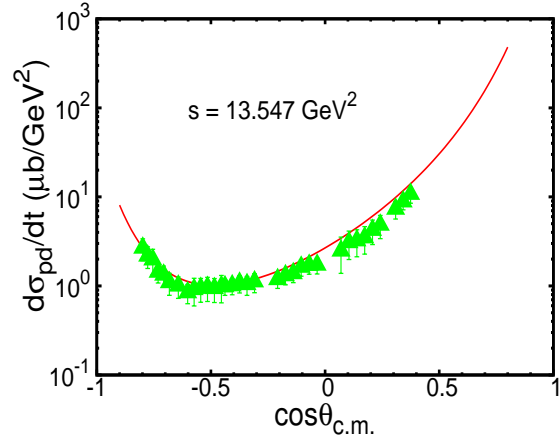}%
}\vfill
\subfloat[\label{pdfit3}] {%
\includegraphics[height = 7 cm, width = 8 cm]{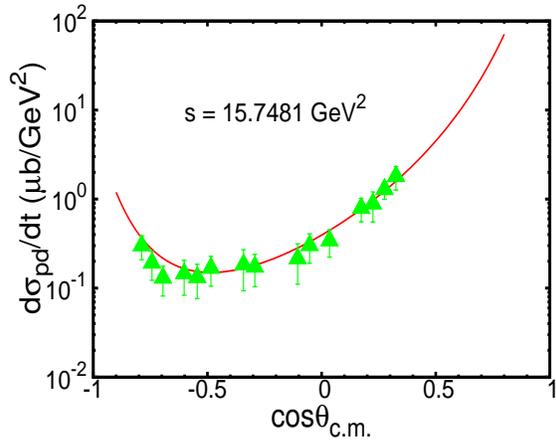}%
}\hfill
\subfloat[\label{pdfit4}]{%
\includegraphics[height = 7 cm, width = 8 cm]{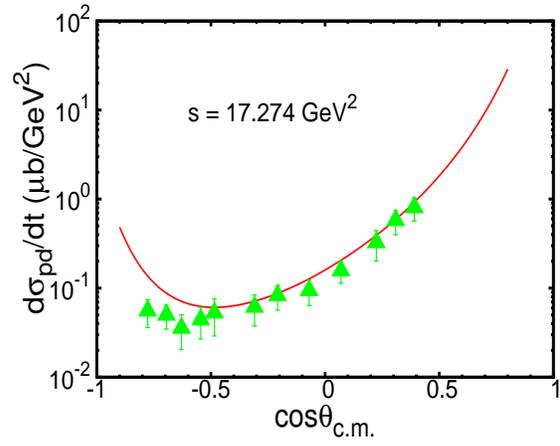}%
}
\captionsetup{format = plain, justification = raggedright}
\caption{Fits of elastic $pd \rightarrow pd$ scattering cross section. The data in (a) are from \cite{800MeV} and 
in (b),(c) and (d) are from \cite{4.50GeV}. The curves are the fits to the  data obtained using Eq. (\ref{pd_paramterization}) with fit parameters from  Table (\ref{table_fitparams}).}
\label{Figpdfits}
\end{figure}

To achieve this,  we  parametrized the existing experimental data on elastic $pd\rightarrow pd$ 
scattering \cite{1.0GeV,582MeV,641.3MeV,800MeV,4.50GeV} which covers the invariant energy range of 
$s \sim$ 9.5 - 17.3 GeV$^2$. 
The following  parametric form is used to fit the $pd \rightarrow pd$ cross section data:
\begin{equation} 
\label{pd_paramterization}
\frac{d\sigma_{pd}}{dt}(s,\cos\theta_{cm}^*) = \frac{1}{(s/10)^{16}}  \frac{ A(s) e^{B(\cos\theta_{cm}^*)} } {(1 - \cos^2\theta_{cm}^*)^3},
 \end{equation}  	
where $A(s) = Ce^{(a_1s + a_2s^2)}$ and  $B(x) =   b x + c x^2$, with the fit parameters given in Table~\ref{table_fitparams}.
The samples of fits obtained for the elastic $pd \rightarrow pd$ hard scattering are presented in Fig. \ref{Figpdfits}. \\ 

\begin{table}[h!]
\centering
\begin{tabular}{| c | c | c | c | c | } % | is the vertical separator, c indicates cell values centered
\hline
  $C$~($\mu$b GeV$^{30}$) & $a_1$~(GeV$^{-2}$) & $a_2$~(GeV$^{-4}$) & b &  c  \\ 
 \hline
  (9.72 $\pm$ 1.33) $\times$ 10$^4$  & -0.98 $\pm$ 0.05  & 0.04 $\pm$ 0.001  &  3.45 $\pm$ 0.02 &  -0.83 $\pm$ 0.05 \\ 
 \hline 
\end{tabular}
\caption{Fit Parameters.}
\label{table_fitparams}
\end{table}

The errors quoted in the table for the fitting parameters result in a  overall error in the $pd\to pd$ cross section 
on the level of 22-37\%.  Note that the  form of the ansatz used in Eq. (\ref{pd_paramterization}) is in agreement with the 
energy and angular dependence following from the quark interchange mechanism of the $pd$ elastic scattering.  As a result
the ansatz is strictly valid for  large center-of-mass angles $|cos(\theta_{cm}^*)| \le 0.6$.  However we extended the 
fitting  procedure beyond this angular range by introducing an additional function $e^{B(\cos\theta_{cm}^*)}$.

\subsubsection{Estimation of the effective charge $Q_{eff}$.}
To calculate the effective quark charge associated with the hard rescattering amplitude, we notice that from Eq. (\ref{simplifiedAmplitude}) it follows that $Q_{eff}$ should satisfy the following relation:
\begin{equation}
 \sum\limits_{i \in p} Q_i\langle d'p' \mid M_{pd,i}\mid dp\rangle     =  Q_{eff}\langle d'p' |M_{pd}\mid dp\rangle,
\end{equation}
 where by $i$ we sum by  the quarks in the proton that were struck by incoming photon.  To use  the above equation one needs a specific model for $pd$  elastic scattering which explicitly uses underlying  quark degrees of the freedom in  $pd$ scattering.  
 For such a model we use the quark-interchange  mechanism~(QIM).  The consideration of a quark-interchange mechanism is justified if one works in the regime in which  the $pd$ elastic scattering exhibits  scaling in agreement with quark counting rule, i.e., $s^{-16}$.  
 
 Similar to Refs.\cite{gdpn,gheppn,gdBB}, within the QIM  $Q_{eff}$ can be estimated using the relation
\begin{equation}
Q_{eff}= \frac{N_{uu}(Q_u) + N_{dd}(Q_d) + N_{ud}(Q_u+Q_d)}{ N_{uu} + N_{dd} + N_{ud}},
\label{QF}
\end{equation}
where $Q_i$ is the charge of the $u$ and $d$ valence quarks  in the proton $p$ and $N_{ii}$ represents 
the number of quark interchanges for $u$ and $d$ flavors   necessary to produce a given helicity $pd$ amplitude. Note that for the particular case of elastic $pd$ scattering $N_{ud} = 0$, and one obtains $Q_{eff} = \frac{1}{3}$.

\subsubsection{Final expression for the differential cross section}
\begin{figure}[ht]
\centering
\includegraphics[width = 12 cm, height = 9 cm]{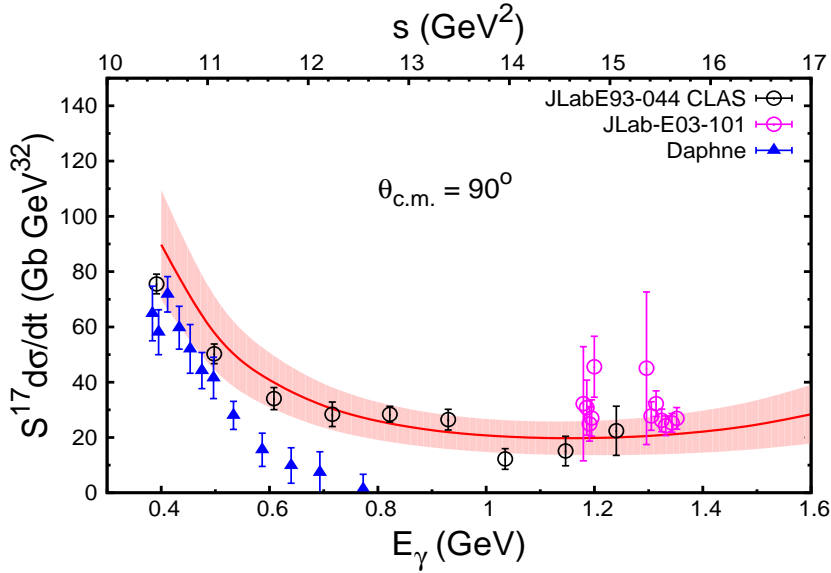}
\captionsetup{format = plain, justification = raggedright}
\caption{Energy dependence of the differential cross section at $\theta_{cm} = 90^0$ scaled by an $s^{17}$ factor. The solid 
curve is the calculation according to Eq. (\ref{diffeq_b}).  The experimental data are from Refs.\cite{DAPHNE,Pomerantz13,Berman}. 
See also the discussion in Ref.\cite{Pomerantz13} on disagreement between DAPHNE and JLAB/CLAS data. }
\label{comp}
\end{figure}

Substituting Eq. (\ref{LCspectralNR}) into Eq. (\ref{differentialCS})  and taking into account the above estimation of 
$Q_{eff}$, we arrive at the final expression for the differential cross section which will be used for 
the numerical estimates:
\begin{equation}
\frac{d\sigma}{dt}(s,t) = \frac{2 \pi^4 \alpha}{3 s'_{^3\text{He}}} \Big(\frac{s'_N}{s'_{^3\text{He}}}\Big) \frac{d\sigma_{pd}}{dt}(s, t_{pd}) \cdot m_N S^{NR}_{^3\text{He}/d}(p_{1z}=0),
\label{diffeq_b}
\end{equation}
where $\alpha = \frac{e^2}{4\pi}$ is the fine structure constant.
For the evaluation of the transition spectral function $S^{NR}_{^3\text{He}/d}$,  we use  the realistic $^3\text{He}$~\cite{Nogga:2002} and deuteron~\cite{Wiringa:1994wb}
wave functions  based on the  V18 potential\cite{Wiringa:1994wb} of $NN$ interaction. 
This yields\cite{eheppnII}  $S^{NR}_{^3\text{He}/d}(p_{1z}=0) = 4.1\times 10^{-4}$~GeV.  For the differential cross section of 
the large center-of-mass  elastic $pd\rightarrow pd$ scattering, $\frac{d\sigma_{pd}}{dt}(s, t_{pd})$, we use Eq. (\ref{pd_paramterization}) which covers the invariant energy range of up to $s=17.3$~GeV$^2$, corresponding to $E_\gamma = 1.67$~GeV for the reaction~(\ref{reaction}). In Fig. \ref{comp} 
we present the comparison of  our calculation of the energy dependence of  the $s^{17}$ scaled differential cross section  
at $\theta_{cm} = 90^0$ with the data of Ref.\cite{Pomerantz13}. The shaded area represents the error due to the above discussed fitting of the elastic $pd\rightarrow pd$ cross sections. 

As the comparison shows,  Eq. (\ref{diffeq_b}) describes surprisingly well the Jefferson Lab data, considering the 
fact that the cross section between $E_{\gamma} = 0.4$~GeV and $E_{\gamma} = 1.3$~GeV drops by a factor of $\sim 4000$.
It is interesting that  the HRM model describes data reasonably well even for the range of 
$E_{\gamma} < 1$~GeV for which the general conditions for the onset of QCD degrees of freedom is not satisfied (see the discussion in Sec. \ref{sectionkin}).  This situation is specific to the HRM model in which there is another scale $t_{pd}$, the invariant momentum transfer in the hard rescattering amplitude. The $-t_{pd} > 1$ GeV$^2$  condition is necessary for the factorization of the  hard scattering kernels from the soft nuclear parts.  It follows from Eq. (\ref{t_pd}) that such a threshold for $t_{pd}$ is reached already for  incoming photon  energies of 0.7~GeV. Considering the photon energies below 0.7 GeV, the qualitative agreement of the HRM model with the data is an indication  of the smooth transition from the hard to the soft regime of the interaction.

\medskip

\begin{figure}[h!]
\centering
\includegraphics[width = 12 cm, height = 9 cm]{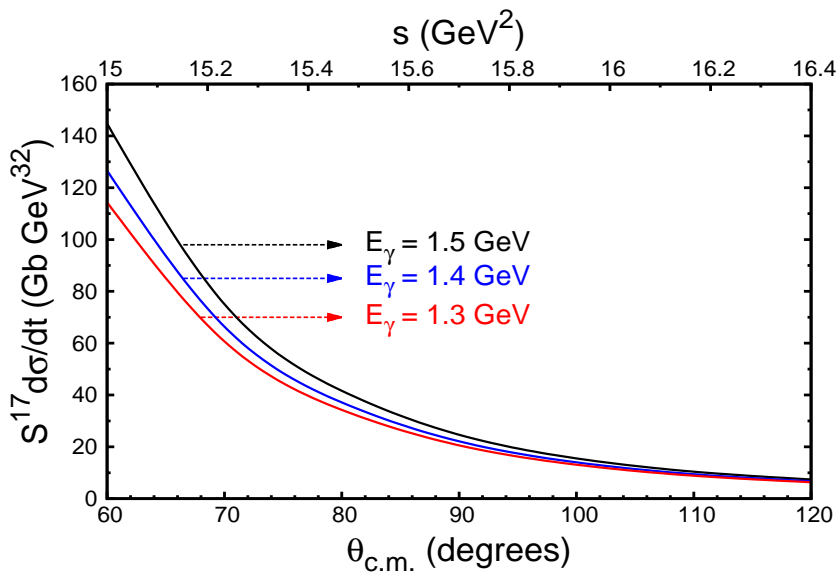}
\captionsetup{format = plain, justification = raggedright}
\caption{The angular dependence  of $s^{17}$ scaled differential cross section for incoming photon energies of 1.3, 1.4, and 1.5~GeV.}
\label{ang_deps}
\end{figure}

The HRM model allows us to also calculate the angular distribution of the differential cross section for fixed values of $s$. In Fig. \ref{ang_deps} we present predictions for angular distribution of the energy scaled differential cross section at largest photon energies for which there are  available data\cite{YI}. 

The interesting feature of the HRM prediction is that, due to the fact that the magnitude of  invariant momentum transfer of the 
reaction~(\ref{reaction}), $t$, is larger than that of   the $pd\rightarrow pd$ scattering, $t_{pd}$ [Eq. (\ref{t_pd})], the 
effective center-of-mass angle   in the latter case, $\theta^*_{cm}< \theta_{cm}$ (see Fig. \ref{eff_cm_angle}); as a result HRM predicts angular distributions that are monotonically decreasing with an increase of $\theta_{cm}$ for up to $\theta_{cm}\approx 120^0$.

\begin{figure}[h!]
\centering
\includegraphics[width = 12 cm, height = 9 cm]{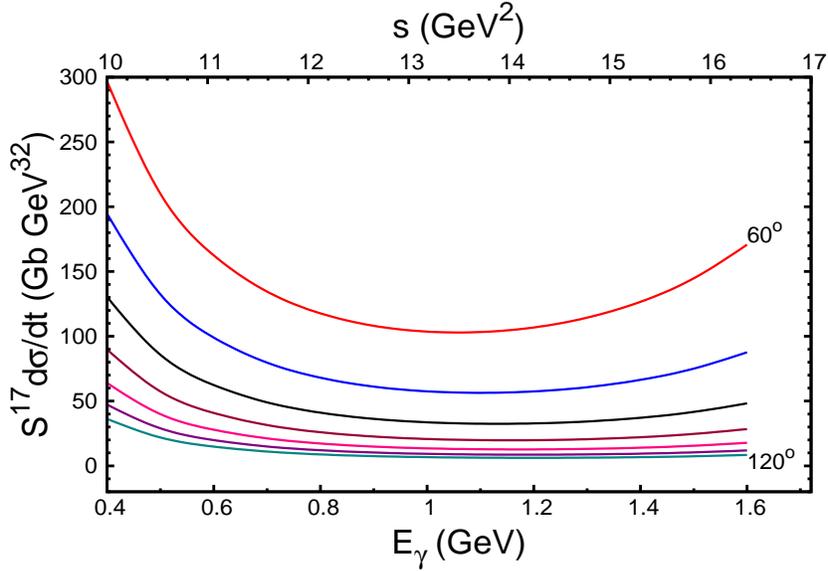}
\captionsetup{format = plain, justification = raggedright}
\caption{Energy dependence of the $s^{17}$ scaled differential cross section  for different values of $\theta_{cm}$.  
The upper curve corresponds to $\theta_{cm} = 60^o$, with the following curves corresponding to increment of the center-of-mass angle by $10^o$. }
\label{diff_angles}
\end{figure}

Finally in Fig. \ref{diff_angles} we present the calculation of the $s^{17}$ scaled differential cross section as a function of incoming photon energy for 
different fixed  and large center-of-mass angles, $\theta_{cm}$.  
Note that in both Figs. \ref{ang_deps} and \ref{diff_angles} the  accuracy of the theoretical predictions is similar to that of the energy dependence at $\theta_{cm}=90^0$ presented in Fig. \ref{comp}.

The possibility of comparing these calculations with the experimental data will allow us
to ascertain the range of  validity of the HRM mechanism.  These comparisons will allow us to identify the minimal momentum transfer in these nuclear reactions for which one observes  the onset of QCD degrees of freedom.

\section{Summary and Outlook}
We extended the consideration of the hard rescattering mechanism of two-body breakup reactions to the 
high energy photodisintegration of the $^3\text{He}$ target to the ($p,d$) pair  at large center-of-mass angles.  
The obtained expression for the cross section does not contain free parameters and is expressed through the effective charge of the constituent quarks  being struck by the incoming photon and interchanged in the final state of the 
process, the  $^3\text{He}\rightarrow pd$ transition spectral function, and the hard elastic $pd\rightarrow pd$ scattering differential cross section.  

For numerical results we estimated the effective quark charge based on the quark-interchange model of $pd\rightarrow pd$ scattering. The transition spectral function is calculated using realistic wave functions of $^3\text{He}$ and the deuteron and  
the $pd\rightarrow pd$ cross section is taken from the experiment.
The calculated differential cross section of reaction~(\ref{reaction}) at $\theta_{cm} = 90^0$ is compared with the recent 
experimental data from Jefferson Lab.  The comparison shows a rather good agreement with the data for the range of photon 
energies $E_{\gamma}\gtrsim 0.7$~GeV.  We also give predictions for angular distribution of the cross section, which reflects 
the special property of HRM in which the magnitude of the invariant momentum transfer entering in the reaction~(\ref{reaction}) 
exceeds the one entering in the hard amplitude of $pd\rightarrow pd$ scattering.  
The possibility of comparing the energy dependence of the cross section for different $\theta_{cm}$ of the $pd$ breakup will allow us to establish the  kinematic boundaries  in which  QCD degrees of freedom are important for the quantitative 
description of the hard $pd$ breakup reactions.

\medskip
\medskip
\noindent {\bf Acknowledgements:} We are thankful to Drs. Y.~Ilieava, E. Piasetzky, and I.~Pomerantz for numerous discussions and comments as well 
as explanation of the experimental data.  This work is supported by a U.S. DOE grant under Contract No. DE-FG02-01ER41172.

\appendix
\renewcommand{\theequation}{A-\arabic{equation}}
\renewcommand\thefigure{ A.\arabic{figure}}
\renewcommand{\thesubsection}{A.\arabic{subsection}}
\setcounter{equation}{0}
\section*{Appendix A}

\subsection{ Nuclear and Nucleonic wave functions}
In this appendix, the details on derivation of the wave functions of the  $^3\text{He}$ nucleus and deuteron are discussed.
We begin  with considering the part   ${\bf A}$  in Eq. (\ref{StartDerivationAmplitude}) related to the $^3\text{He}\rightarrow d$ nuclear transition
\begin{eqnarray} \label{A1}
 &&A_1  = \int (-i) \frac{\chi_{d}^{*^{\lambda_{d}}} \Gamma_{_{DNN}}^{\dagger}}{p_{d}^2 - m_{_d}^2 + i\epsilon} \frac{i(\slashed{p}_{3}+m)}{p_{3}^{2}-m_N^2+i\epsilon} \frac{i(\slashed{p}_{2}+m)}{p_{2}^{2}-m_N^2+i\epsilon} \frac{i(\slashed{p}_{1}+m)}{p_{1}^{2}-m_N^2+i\epsilon}  \nonumber \\
&& \times \qquad  i\Gamma_{_{^3\text{He}}} \chi_{_{^3\text{He}}}^{\lambda_{^3\text{He}}} \frac{1}{2}\frac{dp_{_{2+}}dp_{_{2-}}d^2p_{_{2\perp}}}{(2\pi)^4} \frac{1}{2}\frac{dp_{_{3+}}dp_{_{3-}}d^2p_{_{3\perp}}}{(2\pi)^4},
\end{eqnarray}
where the  $d^4p$  differentials are expressed in terms of the light cone momenta.
The denominators of the propagators in this expression can be expanded as follows:
\begin{equation}
p^2 - m_N^2 + i\epsilon = p_{_+} \Big(p_{_-} - \frac { m_N^2 + p_{_\perp}^2 }{p_+} + i\epsilon'\Big),
\end{equation}
in which using the relations $p_{1-} = p_{^3\text{He}-} - p_{d-}$ and  $p_{d-} = \frac{m_d^2 + p_{d\perp}^2}{p_{d+}}$ one obtains
\begin{equation} \label{denom}
p_1^2 - m_N^2 + i\epsilon  = \frac{1}{p_{_{^3\text{He}+}}} \Big(m_{_{^3\text{He}}}^2 - \frac{m_{_d}^2 + p_{_{d\perp}}^2}{\beta_d} - \frac{m_N^2 + p_{_{1\perp}}^2}{\beta_1}\Big),
\end{equation}
where  $ \beta_1 = \frac{p_{_{1+}}}{p_{_{^3\text{He}+}}}$,  $\beta_d = \frac{p_{_{d+}}}{p_{_{^3\text{He}+}}} $ and $\beta_1 + \beta_d = 1$.
Using the sum rule relation, $ (\slashed{p} + m) = \sum_{\lambda} u(p,\lambda)\bar{u}(p,\lambda)$,
one introduces the light-front wave function of   $^3\text{He}$ (see, e.g., \cite{FS81,pAjj,multisrc}) as follows:
\begin{equation} \label{He3wf}
\Psi_{_{^3\text{He}}}^{^{\lambda_{^3\text{He}}}}(\beta_1,\lambda_1,p_{_{1\perp}},\beta_2,p_{_{2\perp}}\lambda_2,\lambda_3)=\frac{\bar{u}(p_3,\lambda_3)\bar{u}(p_2,\lambda_2)\bar{u}(p_1,\lambda_1)}{\Big(m_{_{^3\text{He}}}^2 - \frac{m_{_d}^2 + p_{d\perp}^2}{\beta_{d}}- \frac{m_N ^2 + p_{{1\perp}}^2 }{\beta_1}\Big)} \Gamma_{_{^3\text{He}}} \chi_{_{^3\text{He}}}^{\lambda_{^3\text{He}}}.
\end{equation}
This wave function gives the probability amplitude of finding 
the  $^3\text{He}$ nucleus with helicity $\lambda_{^3\text{He}}$ consisting of nucleons  with momenta $p_i$  and helicities $\lambda_i$, $i=1,2,3$.  
Using  the above  definition of the $^3\text{He}$ wave function and Eq. (\ref{denom}) in  Eq. (\ref{A1}), one obtains
\begin{eqnarray} \label{FirstSegwithHe3wf}
&& A_1 =  -i  \sum_{\lambda_{d},\lambda_{3},\lambda_{2},\lambda_{1}} \int \frac{\chi_{d}^{*^{\lambda_{d}}} \Gamma_{_{DNN}}^{\dagger}}{p_{d}^2 - m_{_d}^2 + i\epsilon} u(p_3,\lambda_3) u(p_2,\lambda_2) u(p_1,\lambda_1)  \frac{1}{p_{_{2+}}\Big(p_{_{2-}} - \frac{m_N^2 + p_{_{2\perp}}^2}{p_{_{2+}}}\Big)} \nonumber \\
  && \times \frac{\Psi_{_{^3\text{He}}}^{^{\lambda_{^3\text{He}}}}(\beta_1,\lambda_1,p_{_{1\perp}},\beta_2,p_{_{2\perp}}\lambda_2,\lambda_3)}{\beta_1} \frac{1}{2} \frac{dp_{_{2-}}dp_{_{2+}}d^{2}p_{_{2\perp}}}{(2\pi)^4} \frac{1}{2} \frac{d\beta_3}{\beta_3}\frac{d^2p_{_{3\perp}}}{(2\pi)^3},
\end{eqnarray} where $\beta_3 = \frac{p_{3+}}{p_{^3\text{He}+}}$ and the integral over $p_{3-}$ is performed at its pole value: 
\begin{equation}\label{polevalueint_scheme}
\int \frac{dp_-}{p_- - \frac{m_N^2 + p_\perp^2}{p_+} + i\epsilon} = -2\pi i |_{p_- = \frac{m_N^2 + p_\perp^2}{p_+}}.
\end{equation} 
Since $p_2 + p_3 = p_d$, the differentials with respect to $p_2$ can be written as
\begin{equation} \label{p2minustopdminus}
\frac{dp_{_{2-}}dp_{_{2+}}dp_{_{2\perp}}}{(2\pi)^4} = \frac{dp_{_{d-}}dp_{_{d+}}dp_{_{d\perp}}}{(2\pi)^4}.
\end{equation}
Also, the quantity $p_{_{2+}}\big(p_{_{2-}} - \frac{m_N^2 + p_{_{2\perp}}^2 }{p_{_{2+}}}\big)$  in Eq. (\ref{FirstSegwithHe3wf})  
can be represented as:
\begin{equation} \label{p2plusdenom}
p_{_{2+}}\Big(p_{_{2-}} - \frac{p_{_{2\perp}}^2 + m_N^2}{p_{_{2+}}}\Big) =  (1 - \alpha_3)\Big(m_{_{d}}^{2} + p_{_{d\perp}}^{2} - \frac{m_N^2 + p_{_{3\perp}}^{2}}{\alpha_{3}} - \frac{ m_N^2 + p_{_{2\perp}}^{2}}{1- \alpha_{3}}\Big) ,
\end{equation}
where we define $ \alpha_{3} = \frac{p_{_{3+}}}{p_{_{d+}}} \equiv \frac{\beta_3}{\beta_{d}} $ and $ 1 - \alpha_3 \equiv \alpha_{2} = \frac{p_{_{2+}}}{p_{_{d+}}} =  \frac{\beta_2}{\beta_d} \text{ with } \beta_d = \frac{p_{d+}}{p_{^3\text{He}+}} \text{ and } \beta_2 = \frac{p_{2+}}{p_{^3\text{He}+}} $. The quantities $\alpha_{3}$ and $ \alpha_{2} $ represent the fractions of the momentum of the intermediate deuteron carried by the nucleons 3 and 2 respectively. Note that $ \alpha_{3} + \alpha_{2}=1$.

Similar to Eq. (\ref{He3wf}), we introduce light-front   wave function of the deuteron\cite{FS81, pAjj, multisrc}:
\begin{equation} \label{DeuteronWF}
\Psi_{_{d}}^{\lambda_{d}}(\alpha_{3},p_{_{3\perp}},p_{_{d\perp}}) = \frac{\bar{u}(p_2,\lambda_2) \bar{u}(p_3,\lambda_3)}{\Big(m_{_{d}}^{2} + p_{_{d\perp}}^{2} - \frac{m_N^2 + p_{_{3\perp}}^{2}}{\alpha_{3}} - \frac{ m_N^2 + p_{_{2\perp}}^{2}}{1- \alpha_{3}}\Big)} \Gamma_{_{DNN}}\chi_{_d}^{^{\lambda_{d}}},
\end{equation}
which describes the probability amplitude of finding in the $\lambda_{d}$-helicity  deuteron two nucleons with  momenta $p_i$ and  helicities $\lambda_i$, $i=1,2$.
Using Eqs.(\ref{p2plusdenom}) and (\ref{DeuteronWF})   for  Eq. (\ref{FirstSegwithHe3wf}) we obtain
\begin{eqnarray}
& A_1 = & -i  \sum_{\lambda_{d},\lambda_{3},\lambda_{2},\lambda_{1}} \int \frac{\Psi_{_{d}}^{\dagger\lambda_{d}:\lambda_2, \lambda_3}(\alpha_{3},p_{_{3\perp}},p_{_{d\perp}})}{1-\alpha_{3}} u(p_1,\lambda_1) \frac{\Psi_{_{^3\text{He}}}^{^{\lambda_{^3\text{He}}}}(\beta_1,\lambda_1,p_{_{1\perp}},\beta_2,p_{_{2\perp}}\lambda_2,\lambda_3)}{\beta_1} \nonumber \\
&& \times \frac{1}{2} \frac{d\beta_{_d}d^{2}p_{_{d\perp}}}{\beta_{d}(2\pi)^3} \frac{1}{2} \frac{d\beta_3}{\beta_3}\frac{d^2p_{_{3\perp}}}{(2\pi)^3},
\label{A1part}
\end{eqnarray}
where the  $p_{2-}$ integration is performed similar to that of $p_{3-}$ according to Eq. (\ref{polevalueint_scheme}).

\medskip
\medskip

Next, we consider the  second part of the expression $A$ in Eq. (\ref{StartDerivationAmplitude}) related to the transition of the 
deuteron from intermediate to the final state:
\begin{eqnarray} \label{B1part}
& A_2 &=  -\int \chi_{d}^{*\lambda_{d}}(\Gamma_{_{DNN}}^{\dagger}) \frac{i(\slashed{p}_{2f} + m)}{p_{2f}^{2} - m_N^2 + i\epsilon} \frac{i(\slashed{p}_{3}'+m)}
{p_{3}'^{2}-m_N^2+i\epsilon} \frac{i(\slashed{p}_{2}'+m)}{p_{2}'^{2}-m_N^2+i\epsilon} i\Gamma_{_{DNN}} \chi_{d}^{\lambda_{d}} \frac{d^4p_{3}'}{(2\pi)^4} \nonumber \\ 
&&= -\sum_{\substack{\lambda_{2f} \\ \lambda_{2}', \lambda_{3}'}} \int \chi_{d}^{*\lambda_{d}} \Gamma_{_{DNN}}^{\dagger} \frac{u(p_{2f},\lambda_{2f})
\bar{u}(p_{2f},\lambda_{2f})}{p_{2f}^{2} - m_N^2 + i\epsilon}  \frac{u(p_{2}',\lambda_{2}')\bar{u}(p_{2}',\lambda_{2}')}{p_{_{2+}}'
\Big(p_{_{2-}}' - \frac{m_N^2 + p_{2\perp}'^{2}}{p_{_{2+}}'}+i\epsilon\Big)}  \nonumber  \\ 
&& \times \qquad u(p_{3}',\lambda_{3}')\bar{u}(p_{3}',\lambda_{3}') i\Gamma_{_{DNN}} \chi_{d}^{\lambda_{d}} \frac{1}{2} 
\frac{dp_{_{3+}}'}{p_{_{3+}}'}\frac{d^2p_{_{3\perp}}'}{(2\pi)^3},
\end{eqnarray}
where the $dp_{3-}'$ is integrated according to Eq. (\ref{polevalueint_scheme}). To estimate the denominator $p_{2f}^2 - m_N^2 + i\epsilon$ we use the relation
$p_{2f} = p_{df} - p_{3}'$, which allows us to express
\begin{equation} \label{p2fplusdenom}
 p_{2f}^2 - m_N^2 + i\epsilon =   \frac{p_{_{2f+}}}{p_{_{df+}}} \Big( m_d^2 + p_{_{df\perp}}^2 - \frac{m_N^2 + p_{_{3\perp}}'^{2}}{p_{_{3+}}'/p_{_{df+}}} - \frac{m_N^2 + p_{_{2f\perp}}^2}{p_{_{2f+}}/p_{_{df+}}} + i\epsilon \Big).
\end{equation}
Defining $ \frac{p_{_{3+}}'}{p_{_{df+}}}  = \frac{p_{_{3+}}'/p_{_{d+}}}{p_{_{df+}}/p_{_{d+}}} = \frac{\alpha_{3}'}{\gamma_{d}} $ and $\frac{p_{_{2f+}}}{p_{_{df+}}}  = 1 - \frac{p_{_{3+}}'}{p_{_{df+}}} = 1 - \frac{\alpha_{3}'}{\gamma_{d}}$,  the above equation reduces to:
\begin{equation} \label{p2fplusdenomfinal}
 p_{2f}^2 - m_N^2 + i\epsilon = (1 - \frac{\alpha_{3}'}{\gamma_{d}})\Big( m_d^2 + p_{_{df\perp}}^2 - \frac{m_N^2 + p_{_{3\perp}}'^{2}}
 {\alpha_{3}'/\gamma_{d}} - \frac{m_N^2 + p_{_{2f\perp}}^2}{1 - \alpha_{3}'/\gamma_{d}}+ i\epsilon \Big),
\end{equation}
where the quantity $\gamma_{d} = \frac{p_{_{df+}}}{p_{_{d+}}} $ is the fraction of the momentum of the intermediate deuteron carried by the final deuteron.
Using  Eqs.(\ref{DeuteronWF})  and (\ref{p2fplusdenomfinal}), we rewrite Eq. (\ref{B1part}) as follows:
\begin{eqnarray}
& A_2 &= \sum_{\substack{\lambda_{df},\lambda_{2f} \\ \lambda_{2}', \lambda_{3}'}} \int \frac{{\Psi_{d}^{\dagger \lambda_{df}:\lambda_{3}', \lambda_{2f}}}(\alpha_{2f}/\gamma_{d},p_{2\perp},\alpha_{3}'/\gamma_{d},p_{3\perp}')}{1-\alpha_{3}'/\gamma_{d}} \bar{u}(p_{2f},\lambda_{2f}) u(p_{2}',\lambda_{2}')  \nonumber \\
&& \times \qquad \frac{\Psi_d^{\lambda_{d}:\lambda_{2}', \lambda_{3}'}(\alpha_{3}',p_{d\perp},p_{3\perp}')}{1-\alpha_{3}'} 
\frac{1}{2} \frac{d\alpha_{3}'}{\alpha_{3}'}	\frac{d^2p_{3\perp}'}{(2\pi)^3}.
\label{A2part}
\end{eqnarray}

\medskip
Now we consider the $N1$ part of the amplitude in Eq. (\ref{StartDerivationAmplitude}), which describes the transition of the nucleon with momentum $p_1$ to 
the final nucleon with momentum $p_{1f}$.  Using on-shell  sum-rule relations for the numerators of the quark propagators for the $N1$ part, one has
\begin{eqnarray} \label{N1start}
&& N1 =\sum_{\substack{\lambda_1 \\ \eta_{1f},\eta_1 }}\int \bar{u}(p_{1f}, \lambda_{1f}) (-i)\Gamma_{n1f}^{\dagger} 
\frac{u_q(p_{1f}-k_1,\eta_{1f})\bar{u}_q(p_{1f}-k_1,\eta_{1f})}{(p_{1f} - k_1)^{2} - m_q^{2} + i\epsilon} \Big[-igT_{c}^{\beta} \gamma_{\mu}\Big]  \nonumber \\ 
&&\times \frac{\Psi_s(k_1)\bar{\Psi}_s(k_1)}{k_{1}^{2} - m_s^{2} + i\epsilon} 
  \frac{u_q(p_1-k_1,\eta_1)\bar{u}_q(p_1-k_1,\eta_1)}{(p_{1} - k_1)^{2} - m_q^{2} + i\epsilon} i\Gamma_{n1} 
  u(p_1,\lambda_1)\frac{1}{2}\dfrac{dk_{1+} dk_{1-} d^2k_{1\perp}}{(2\pi)^4}, 
\end{eqnarray}
where we sum over the initial helicity ($\eta_1$) of the quark before being struck by the incoming photon and the final helicity ($\eta_{1f}$) of the quark that recombines to form the final state proton.
In Eq. (\ref{N1start}), we can expand the denominators of the propagators as follows:
\begin{eqnarray} \label{propagator_denom_N1}
&& (p_{1f} - k_{1})^2 - m_q^2 + i\epsilon= (1 - x_{s1}) \Big( m_N^2 + p_{1f\perp}^2 - \frac{m_{s}^2 + k_{1\perp}^2}{x_{s1}} - \frac{m_q^2 + (p_{1f} - k_1)_{\perp}^2 }{1-x_{s1}} + i\epsilon\Big) \nonumber \\
&& (p_1 - k_1)^2 - m_q^2 + i\epsilon= (1 - x_1)\Big( m_N^2 + p_{1\perp}^2 - \frac{m_{s}^2 + k_{1\perp}^2}{x_{1}} - \frac{m_q^2 + (p_1 - k_1)_{\perp}^2}{1-x_{1}}+ i\epsilon \Big),
\nonumber \\
& & 
\end{eqnarray}
where $x_{s1}=\frac{k_{1+}}{p_{1f+}} \text{ and } x_1 = \frac{k_{1+}}{p_{1+}}$ along with $k_{1-} = \frac{m_{s}^{2} + k_{1\perp}^2}{k_{1+}} $ and $ p_{1f-} = \frac{m_N^2 + p_{1f\perp}^2}{p_{1f+}}$. Here  $x_1 (x_{s1})$ is interpreted as the momentum fraction of the initial (final) nucleon ``1" carried  by the spectator quark system. 
Performing the  $dk_{1-}$ integration at the $k_{1-}$ pole value of the spectator system  allows us to introduce a single quark wave function of the nucleon in the  form
\begin{equation} \label{QuarkWF}
\Psi_n^{\lambda;\eta}(X, k_\perp, p_\perp) = \frac{\bar{u}_{q}(p-k,\eta) \bar{\Psi}_s(k)}{m_N^2 + p_\perp^2 - \frac{m_s^2 + k_{\perp}^2}{X} - \frac{m_q^2 +  (p - k)_{\perp}^2 }{1-X}}\Gamma_n u(p,\lambda),
\end{equation}
which describes the probability amplitude of finding  a quark with helicity $\eta$ and momentum fraction $1-x$ in the $\lambda$-helicity nucleon with momentum $p$.
With this definition of quark wave function of the nucleon one obtains for the $N1$ part
\begin{eqnarray} \label{N1withQuarkWF}
&N1& = i\sum_{\substack{ \lambda_1 \\ \eta_{1f}, \eta_1}} \int	\frac{\Psi_{n1f}^{\dagger \lambda_{1f};\eta_{1f}} (x_{s1},k_{1\perp}, p_{1f\perp})}{1 - x_{s1}} 
\bar{u}_q(p_{1f} - k_1,\eta_{1f}) u_q(p_1 - k_1,\eta_1)  \nonumber \\ 
&& \times \Big[ -igT_{c}^\beta \gamma_{\mu}\Big]\frac{\Psi_{n1}^{\lambda_1;\eta_1}(x_1, k_{1\perp},p_{1\perp}) }{1 - x_1} \frac{1}{2}\dfrac{dx_1}{x_1} \dfrac{d^2k_{1\perp}}{(2\pi)^3}.
\end{eqnarray}
Performing  very similar calculations for the $N2$ part of  Eq. (\ref{StartDerivationAmplitude}), one obtains:
\begin{eqnarray} \label{N2}
&N2&=-\sum_{\substack{\lambda_{2f},\lambda_{2}' \\ \eta_{2f}, \eta_{2}'}}\int \frac{\Psi_{n2f}^{\dagger \lambda_{2f};\eta_{2f}}(x_{s2},k_{2\perp}, p_{2f\perp})}{1-x_{s2}} \bar{u}_q(p_{2f}-k_2,\eta_{2f})u_q(p_{2}'-k_2,\eta_{2}')  \nonumber \\
&& \times \frac{\Psi_{n2'}^{\lambda_{2}';\eta_{2}'}(x_{2}',k_{2\perp}, p_{2\perp}')}{1-x_{2}'} 
 \frac{1}{2} \dfrac{dx_{2}'}{x_{2}'} \dfrac{d^2k_{2\perp}}{(2\pi)^3},
\end{eqnarray}
where $x_{s2} = \frac{k_{2+}}{p_{2f+}}$ and $x_{2}' = \frac{k_{2+}}{p_{2+}'}$.

\medskip
Substituting  now Eqs. (\ref{A1part}), (\ref{A2part}), (\ref{N1withQuarkWF}), and (\ref{N2}) into Eq. (\ref{StartDerivationAmplitude}), one obtains
\begin{eqnarray} \label{amp_beforePhotonQuarkInteraction}
&&\mathcal{M}^{\lambda_{df},\lambda_{1f};\lambda_{^3\text{He}},h}=  \sum_{\substack{(\lambda_{2f})(\lambda_{2}',\lambda_{3}') (\lambda_{d}) \\ (\lambda_1,\lambda_2,\lambda_3) \\ (\eta_{1f}, \eta_{2f}) (\eta_1, \eta_{2}' ) }}  \int  \frac{{\Psi_{d}^{\dagger \lambda_{df}:\lambda_{3}', \lambda_{2f}}}
(\alpha_{2f}/\gamma_{d},p_{2\perp},\alpha_{3}'/\gamma_{d},p_{3\perp}')}{1-\alpha_{3}'/\gamma_{d}} \Bigg
\{ \frac{\Psi_{n2f}^{\dagger \lambda_{2f};\eta_{2f}}(x_{s2}, k_{2\perp}, p_{2f\perp})}{1-x_{s2}}  \nonumber \\
&& \times \bar{u}_q(p_{2f}-k_2, \eta_{2f}) [-igT_c^\alpha \gamma_\nu]  \Big[\frac{i(\slashed{p}_1 + \slashed{q} - \slashed{p_1}+m_q)}
{(p_1 - k_1 + q)^2 -m_q^2 + i\epsilon} \Big] [-ie\epsilon^\mu \gamma\mu] u_q(p_1 - k_1, \eta_1)  \nonumber \\
&& \times \frac{\Psi_{n1}^{\lambda_1;\eta_1}(x_1, k_{1\perp},p_{1\perp}) }{1 - x_1} \Bigg\}_{1} 
 \Bigg\{ \frac{\Psi_{n1f}^{\dagger \lambda_{1f};\eta_{1f}} (x_{s1},k_{1\perp}, p_{1f\perp})}{1 - x_{s1}} \bar{u}_q(p_{1f}-k_1, \eta_{1f}) 
 [-igT_c^\beta\gamma_\mu] u_q(p_{2}'-k_2,\eta_{2}')   \nonumber \\
 && \times
 \frac{\Psi_{n2'}^{\lambda_{2}';\eta_{2}'}(x_{2}' ,k_{2\perp}, p_{2\perp}')}{1-x_{2}'} \Bigg\}_{2} G^{\mu\nu}(r)  
 \frac{\Psi_d^{\lambda_{d}:\lambda_{2}', \lambda_{3}'}(\alpha_{3}',p_{d\perp},p_{3\perp}')}{1-\alpha_{3}'} \frac{\Psi_{_{d}}^{\dagger \lambda_{d}:\lambda_{2}, \lambda_{3}}(\alpha_{3},p_{_{3\perp}},p_{_{d\perp}})}{(1-\alpha_{3})}  \nonumber \\
&& \times \frac{\Psi_{_{^3\text{He}}}^{^{\lambda_{^3\text{He}}}}(\beta_1,\lambda_1,p_{_{1\perp}},\beta_2,p_{_{2\perp}}\lambda_2,\lambda_3)}{\beta_1}  \frac{d \beta_{d}}{\beta_d} \frac{d^2p_{d\perp}}{2(2\pi)^3} \frac{d \beta_{3}}{\beta_3} 
  \frac{d^2p_{3\perp}}{2(2\pi)^3}
\frac{d \alpha_{3}'}{\alpha_{3}'} \frac{d^2p_{3\perp}'}{2(2\pi)^3}  \frac{dx_1}{x_1}\frac{d^2k_{1\perp}}{2(2\pi)^3} \frac{dx_{2}'}{x_{2}'}\frac{d^2k_{2\perp}}{2(2\pi)^3}.
\end{eqnarray}

\subsection{ Hard scattering kernel}
In Eq. (\ref{amp_beforePhotonQuarkInteraction}), the  expression in  $ \{ \}_{1} \{ \}_{2} G^{\mu\nu}(r) $ describes 
the hard  photon-quark interaction followed by a quark interchange through the gluon exchange.  

\subsubsection{Propagator of the struck quark}
We analyze first the propagator of the struck quark, $ \frac{i(\slashed{p}_1 + \slashed{q} - \slashed{k_1}+m_q)}{(p_1 - k_1 + q)^2 -m_q^2 + i\epsilon}$. 

Using the definition of the reference frame from Eq. (\ref{InitialMomentaofPhotonHelium}) and momentum fraction definitions
 $ \beta_1 = \frac{p_{1+}}{p_{^3\text{He}+}} = \frac{p_{1+}}{\sqrt{s'_{^3\text{He}}}} $ and $x_1 = \frac{k_{1+}}{p_{1+}} $, 
 one can isolate the pole term in the denominator of the struck quark propagator as follows:
\begin{eqnarray} \label{photon_quark_interaction_denom}
&&(p_1 - k_1 + q)^2 - m_q^2 + i\epsilon = (p_{1+} - p_{1+} x_1)(p_{1-} - k_{1-} + q_-) - (p_{1\perp} - k_{1\perp})^2 - m_q^2 + i\epsilon \nonumber \\
%&& =p_{1+}(1 - x_1)\Big( \frac{m_N^2 + p_{1\perp}^2}{p_{1+}} - \frac{m_{s}^{2} + k_{1\perp}^{2}}{k_{1+}} + \sqrt{s'_{^3\text{He}}}~\Big) - (p_{1\perp} - k_{1\perp})^2 - m_{q}^{2} + i\epsilon \nonumber \\
% && =\beta_1 \sqrt{s'_{^3\text{He}}} ( 1 - x_1) \Big( \frac{m_N^2 + p_{1\perp}^2}{\beta_1 \sqrt{s'_{^3\text{He}}}} - \frac{m_{s}^{2} + k_{1\perp}^{2}}{x_1 \beta_1 \sqrt{s'_{^3\text{He}}}} + \sqrt{s'_{^3\text{He}}}~\Big) - (p_{1\perp} - k_{1\perp})^2 - m_{q}^{2} + i\epsilon \nonumber \\
&& = s'_{^3\text{He}} (1 - x_1)\Big( \frac{m_N^2 + p_{1\perp}^2}{s'_{^3\text{He}}} - \frac{m_{s}^{2} + k_{1\perp}^2}{x_1 s'_{^3\text{He}}} + \beta_1 - \frac{ m_{q}^{2} + (p_{1\perp} - k_{1\perp})^2 }{s'_{^3\text{He}}(1-x_1)} \Big) + i\epsilon \nonumber \\
&&= s'_{^3\text{He}} (1 - x_1)(\beta_1 - \beta_s + i\epsilon) ,
\end{eqnarray}
where $ \beta_s = -\frac{1}{s'_{^3\text{He}}}(m_N^2 + p_{1\perp}^2 - \frac{m_s^2 + k_{1\perp}^2}{x_1} - \frac{m_q^2 + (p_{1\perp} - k_{1\perp})^2 }{1-x_1})$. 
Using the sum rule relation [$\slashed{p}+m = \sum\limits_\lambda u(p,\lambda)\bar u(p,\lambda)$] for the numerator of the struck quark propagator together with 
Eq. (\ref{photon_quark_interaction_denom}), one can rewrite Eq. (\ref{amp_beforePhotonQuarkInteraction}) as follows:
\begin{eqnarray} \label{amplitudeBeforePhotonPolarization}
&&\mathcal{M}^{\lambda_{df},\lambda_{1f};\lambda_{^3\text{He}},h}= \sum_{\substack{(\lambda_{2f})(\lambda_{2}',\lambda_{3}') (\lambda_{d}) \\ (\lambda_1,\lambda_2,\lambda_3) \\ (\eta_{1f}, \eta_{2f}) (\eta_1, \eta_{2}' ) (\eta_{q1})}} 
 \int  \frac{{\Psi_{d}^{\dagger \lambda_{df}:\lambda_{3}', \lambda_{2f}}}(\alpha_{2f}/\gamma_{d},p_{2\perp},\alpha_{3}'/\gamma_{d},p_{3\perp}')}{1-\alpha_{3}'/\gamma_{d}}  \nonumber \\ 
&&\times \Bigg\{ \frac{\Psi_{n2f}^{\dagger \lambda_{2f};\eta_{2f}}(x_{s2} ,k_{2\perp},p_{2f\perp})}{1-x_{s2}} \bar{u}_q(p_{2f}-k_2, \eta_{2f})
 [-igT_c^\alpha \gamma_\nu] 
 \Big[\frac{u_q(p_1 + q- k_1,\eta_{q1}) \bar{u}_q(p_1 + q - k_1,\eta_{q1}) }{s'_{^3\text{He}} (1 - x_1)(\beta_1 - \beta_s + i\epsilon)} \Big]   \nonumber \\
 && \times [-ie\epsilon^\mu \gamma_\mu] u_q(p_1 - k_1, \eta_1)  
\frac{\Psi_{n1}^{\lambda_1;\eta_1}(x_1, k_{1\perp},p_{1\perp}) }{1 - x_1} \Bigg\}_{1} \Bigg\{ \frac{\Psi_{n1f}^{\dagger \lambda_{1f};\eta_{1f}} (x_{s1},k_{1\perp}, p_{1f\perp})}{1 - x_{s1}} \bar{u}_q(p_{1f}-k_1, \eta_{1f})  \nonumber \\
&& \times [-igT_c^\beta\gamma_\mu] 
u_q(p_{2}'-k_2,\eta_{2}')
\frac{\Psi_{n2'}^{\lambda_{2}';\eta_{2}'}(x_{2}' ,k_{2\perp},p_{2\perp}')}{1-x_{2}'} \Bigg\}_{2} G^{\mu\nu}(r) \frac{\Psi_d^{\lambda_{d}:\lambda_{2}', \lambda_{3}'}(\alpha_{3}',p_{d\perp},p_{3\perp}')}{1-\alpha_{3}'}   \nonumber \\
&& \times\frac{\Psi_{_{d}}^{\dagger \lambda_{d}:\lambda_{2}, \lambda_{3}}(\alpha_{3},p_{_{3\perp}},p_{_{d\perp}})}{(1-\alpha_{3})} \frac{\Psi_{_{^3\text{He}}}^{^{\lambda_{^3\text{He}}}}(\beta_1,\lambda_1,p_{_{1\perp}},\beta_2,p_{_{2\perp}}\lambda_2,\lambda_3)}{\beta_1}  \frac{d \beta_{d}}{\beta_d} \frac{d^2p_{d\perp}}{2(2\pi)^3} \frac{d \beta_{3}}{\beta_3} \frac{d^2p_{3\perp}}{2(2\pi)^3}\frac{d \alpha_{3}'}{\alpha_{3}'} \frac{d^2p_{3\perp}'}{2(2\pi)^3}  \nonumber \\
&& \times \frac{dx_1}{x_1}\frac{d^2k_{1\perp}}{2(2\pi)^3} \frac{dx_{2}'}{x_{2}'}\frac{d^2k_{2\perp}}{2(2\pi)^3}.
\end{eqnarray}
Note that the  above used sum rule for the numerator of the struck quark propagator is valid for on-shell spinors only.  Our use of this sum rule 
is justified based on the use of the peaking approximation in evaluating  Eq. (\ref{amplitudeBeforePhotonPolarization}), in which 
the denominator of the struck quark is estimated at its pole value.

\subsubsection{ Photon quark interaction}
We now consider the term:
\begin{equation}
\bar{u}_q(p_1-k_1+q,\eta_{q1})[-ie\epsilon_h^\mu \gamma^\mu]u_q(p_1-k_1,\eta_1), 
\label{ugu}
\end{equation}
where the incoming photon with helicity $h$ is described by polarization vectors  $\epsilon_{R/L} = \mp\sqrt{\frac{1}{2}} (\epsilon_1 \pm i\epsilon_2)$ for 
$h=1/(-1)$ respectively. Here $\epsilon_1 \equiv (1,0,0)$  and   $\epsilon_2 \equiv (0,1,0)$.
Using these definitions we express
\begin{equation}
 -\epsilon_h^\mu \gamma^\mu  =  \epsilon^\perp \gamma^\perp = -\epsilon_R \gamma_{L} + \epsilon_L \gamma_{R},
\label{epsgamma}
\end{equation}
where $\gamma_{R/L} = \frac{\gamma_x \pm i\gamma_y}{\sqrt{2}}$. 
%We choose the standard definition of $\gamma$ matrices resulting in:
%\begin{equation}
%\gamma_R = \sqrt{2} \begin{pmatrix}
%0 & 0 & 0 & 1 \\
%0 & 0 & 0 & 0 \\
%0 & -1 & 0 & 0 \\
%0 & 0 & 0 & 0
%\end{pmatrix} ,
%\text{\qquad} \gamma_L = \sqrt{2} \begin{pmatrix}
%0 & 0 & 0 & 0 \\
%0 & 0 & 1 & 0 \\
%0 & 0 & 0 & 0 \\
%-1 & 0 & 0 & 0
%\end{pmatrix} .
%\end{equation}
We also resolve  the spinor of the quark with spin $\alpha$ to the  $\pm$ helicity states as follows:
\begin{equation}
u(p,\alpha) = u^{+}(p,\alpha) + u^{-}(p,\alpha)= \frac{1}{2}(1 + \gamma^5) u(p,\alpha) + \frac{1}{2}(1 - \gamma^5) u(p,\alpha).
\end{equation}
%resulting in
%\begin{eqnarray}
%&& u^{+}(p,1/2) =\frac{\sqrt{E}}{2} \begin{pmatrix}
%1 + \frac{p_z}{E} \\
%\frac{p_R}{E} \\
%1 + \frac{p_z}{E} \\
%\frac{p_R}{E}
%\end{pmatrix} ,
%u^{-}(p,1/2) = \frac{\sqrt{E}}{2} \begin{pmatrix}
%1 - \frac{p_z}{E} \\
%-\frac{p_R}{E} \\
%-(1 - \frac{p_z}{E}) \\
%\frac{p_R}{E} \end{pmatrix} \nonumber \\
%&& u^{+}(p,-1/2) =\frac{\sqrt{E}}{2} \begin{pmatrix}
%\frac{p_L}{E} \\
%1 - \frac{p_z}{E} \\
%\frac{p_L}{E}\\
%1 - \frac{p_z}{E} 
%\end{pmatrix} ,
%u^{-}(p,-1/2) =\frac{\sqrt{E}}{2} \begin{pmatrix}
%\frac{-p_L}{E} \\
%1 + \frac{p_z}{E} \\
%\frac{p_L}{E}\\
%-(1 + \frac{p_z}{E}) 
%\end{pmatrix},
%\end{eqnarray}
%where $p_L = p_x - ip_y \text{ and } p_R = p_x + ip_y$.
Finally, in the reference frame of  Eq. (\ref{InitialMomentaofPhotonHelium}) the light-cone four-momenta $(p_+,p_-,p_\perp)$ of the initial and final quarks 
in Eq. (\ref{ugu}),  in the massless limit, are
\begin{eqnarray} \label{momenta_of_Quarks}
 &&\text{Initial momentum: }p_1 - k_1 =\Big(\beta_1(1 - x_1)\sqrt{s'_{^3\text{He}}}, 0,0 \Big), \nonumber \\
 &&\text{ Final momentum: }p_1 - k_1 + q = \big( \beta_1 (1 - x_1)\sqrt{s'_{^3\text{He}}}, \sqrt{s'_{^3\text{He}}},0),
\end{eqnarray}
where we use the the relations $q_+ = 0$, $p_{1+} = \beta_1 p_{^3\text{He}+}$, and $k_{1+} = x_1 p_{1+}$. Because of the finite $\beta_1\sim\frac{1}{3}$ and small 
$x_1\ll 1$ entering in the amplitude (see Appendix \ref{PeakApp}) one also neglects the {``-"} component of 
the initial quark: $(p_1-k_1)_- \approx \frac{ (p_1-k_1)_{\perp}^2 + m_q^2}{\beta_1(1-x_1)\sqrt{s^\prime_{^3\text{He}}}}\sim 0$.
% one neglected the finite transverse momentum of the initial quark as well as $k_{1-} = \frac{m_s^2 + k_{1\perp}^2}{k_{1+}} \sim 0$. We also used 
%the relation:  

 %with (Eq. (\ref{MomentumDefined_newframe})) and $p_{1+} = \beta_1 p_{^3\text{He}+}, p_{^3\text{He}+} = \sqrt{s'_{^3\text{He}}},  p_{1-} = \frac{m_N^2 + p_{1\perp}^2}{p_{1+}} \sim 0, k_{1+} = x_1 p_{1+}, kthe initial and final momenta of the knocked out quark in the light cone coordinates can be shown to be equal to:
%Also since $p_{\pm} = E \pm p_{z}$,  the initial and final momenta of the knocked out quark in the cartesian coordinates will be equal to:
%\begin{align} \label{momenta_KO_Quark_cartesian}
% &\text{Initial Momentum: }p_1 - k_1 =\Big(\frac{\beta_1(1 - x_1)\sqrt{s'_{^3\text{He}}}}{2}, 0,0 ,\frac{\beta_1(1 - x_1)\sqrt{s'_{^3\text{He}}}}{2}\Big), \nonumber \\
% &\text{ Final Momentum: }p_1 - k_1 + q = \Big(\frac{(1 -  \beta_1(1 - x_1))\sqrt{s'_{^3\text{He}}} }{2}, 0,0,- \frac{(1 -  \beta_1(1 - x_1))\sqrt{s'_{^3\text{He}}} }{2}\Big).
%\end{align}
Using Eq. (\ref{momenta_of_Quarks}) and the above definitions of photon polarization, $\gamma$ matrices, and quark helicity states, one obtains that in 
the quark massless limit the only nonvanishing matrix elements of $\bar u \gamma_\pm u$ are
\begin{eqnarray} \label{matrixelements}
&&\bar{u}^{-}_q(p_1-k_1+q,-\frac{1}{2}) \gamma_{+} u^{-}_q(p_1-k_1,-\frac{1}{2}) = -2\sqrt{2E_1E_2} \nonumber \\
&&\bar{u}^{+}_q(p_1-k_1+q, \frac{1}{2}) \gamma_{-} u^{+}_q(p_1-k_1, \frac{1}{2}) = 2\sqrt{2E_1E_2}, 
\end{eqnarray}
where $E_1 = \beta_1(1 - x_1)\frac{\sqrt{s'_{^3\text{He}}}}{2} \text{ and } E_2 = [1 -  \beta_1(1 - x_1)]\frac{\sqrt{s'_{^3\text{He}}}} {2}$ are the initial and final energies of the struck quark respectively.

Using the above relations for Eq. (\ref{ugu}) one obtains  
\begin{equation} \label{matrix_elems}
\bar{u}_q (p_1 - k_1 + q,\eta_{q1}) [ie\epsilon_h^\perp \gamma^\perp]u_q(p_1 - k_1, \eta_{1}) = ie Q_i2 \sqrt{2E_1E_2}(-h) \delta^{\eta_{q1} h} \delta^{\eta_1 h},
\end{equation}
where $Q_i$ is the charge of the struck quark in units of $e$. The above result indicates that incoming  $h$- helicity  photon selects the quark with the same 
helicity ($h=\eta_1$) conserving it during the interaction ($h = \eta_{q_1}$).

\subsection{ Peaking approximation}
\label{PeakApp}

We now consider the $d\beta_d$ integration in Eq. (\ref{amplitudeBeforePhotonPolarization}), noticing that  $d\beta_d = d\beta_1$ and 
separating the pole and principal value parts in the propagator of the struck quark as follows: 
\begin{equation} \label{IntegralWithPVpart_again}
 \frac{1}{\beta_1 -{\beta_s} + i\epsilon} = -i\pi \delta(\beta_1-{\beta_s}) + \text{P.V.} \int \frac{d\beta_1}{\beta_1 - {\beta_s}}.
\end{equation}
Furthermore, we neglect by P.V. part of the propagator since its contribution comes from the high momentum part of the nuclear wave function, $p_1\sim \sqrt{s^\prime_{^3\text{He}}}$, which is strongly suppressed\cite{gdpn}. The integration with the pole part of the propagator will 
fix the value of $\beta_1 = \beta_s$, and the latter in the massless quark limit and negligible transverse component of $\vec p_1$ can be 
expressed as follows:
\begin{equation} \label{beta_s_lim}
\beta_s = \frac{1}{s'_{^3\text{He}}}\Big[\frac{m_s^2(1-x_1) + k_{1\perp}^2}{x_1(1-x_1)} - m_N^2 \Big].
\end{equation}

Now, using the fact that $^3\text{He}$ wave function strongly peaks at $ \beta_1 = \frac{1}{3}$, one can estimate the ``peaking" value of 
the amplitude in Eq. (\ref{amplitudeBeforePhotonPolarization})  taking $\beta_s= \frac{1}{3}$.
The latter condition results in  $x_1 \rightarrow \frac{3(m_s^2 + k_{1\perp}^2)}{s'_{^3\text{He}}} \sim  0$, since $s'_{^3\text{He}}$ is very large in comparison with the transverse momentum $k_{1\perp}$ of the spectator system. This allows us to approximate  $(1-x_1) \approx 1$. With these approximations, one finds that
\begin{eqnarray}\label{stuck_quark_energy_simplified}
&&E_1 = \beta_1 (1-x_1)\frac{\sqrt{s'_{^3\text{He}}}}{2} = \frac{1}{3}\frac{\sqrt{s'_{^3\text{He}}}}{2}  \nonumber \\ 
&&E_2=\Big(1-\beta_1(1-x_1)\Big)\frac{\sqrt{s'_{^3\text{He}}}}{2}= \frac{2}{3}\frac{\sqrt{s'_{^3\text{He}}}}{2}.
\end{eqnarray}
Using Eq. (\ref{stuck_quark_energy_simplified}) in Eq. (\ref{matrix_elems}) and setting $\beta_1 = 1/3$ everywhere for 
Eq. (\ref{amplitudeBeforePhotonPolarization}) one obtains
\begin{eqnarray} \label{AmpToIntroducePDamp}
&& \mathcal{M}^{\lambda_{df},\lambda_{1f};\lambda_{^3\text{He}},h}=  \frac{3}{4} (-h)\frac{1}{\sqrt{s'_{^3\text{He}}}} \sum_{i}eQ_i
 \sum_{\substack{(\lambda_{2f})(\lambda_{2}',\lambda_{3}') (\lambda_{d}) \\ (\lambda_1,\lambda_2,\lambda_3) \\ (\eta_{1f}, \eta_{2f}) (\eta_{2}' ) }} \int  \frac{{\Psi_{d}^{\dagger \lambda_{df}:\lambda_{3}',\lambda_{2f}}}(\alpha_{2f}/\gamma_{d},p_{2\perp},\alpha_{3}'/\gamma_{d},p_{3\perp}')}{1-\alpha_{3}'/\gamma_{d}}  \nonumber \\
 && \times \qquad \Bigg\{ \frac{\Psi_{n2f}^{\dagger \lambda_{2f};\eta_{2f}}(x_{s2},p_{2f\perp},k_{2\perp})}{1-x_{s2}}  
\bar{u}_q(p_{2f}-k_2, \eta_{2f})
[-igT_c^\alpha \gamma_\nu] 
 \Big[u_q(p_1+q-k_1,h)   \Big]  \nonumber \\
&& \times \frac{\Psi_{n1}^{\lambda_1;h}(x_1, k_{1\perp},p_{1\perp}) }{1 - x_1} \Bigg\}_{1}
 \Bigg\{ \frac{\Psi_{n1f}^{\dagger \lambda_{1f};\eta_{1f}} (x_{s1},k_{1\perp}, p_{1f\perp})}{1 - x_{s1}}
 \bar{u}_q(p_{1f}-k_1, \eta_{1f}) [-igT_c^\beta\gamma_\mu] u_q(p_{2}'-k_2,\eta_{2}')  \nonumber \\
&& \times
 \frac{\Psi_{n2'}^{\lambda_{2}';\eta_{2}'}(x_{2}',p_{2\perp}',k_{2\perp})}{1-x_{2}'} \Bigg\}_{2} G^{\mu\nu}(r) 
 \frac{\Psi_d^{\lambda_{d}:\lambda_{2}',\lambda_{3}'}(\alpha_{3}',p_{d\perp},p_{3\perp}')}{1-\alpha_{3}'} 
 \frac{\Psi_{_{d}}^{\dagger \lambda_{d}:\lambda_{2},\lambda_{3}}(\alpha_{3},p_{_{3\perp}},p_{_{d\perp}})}{1-\alpha_{3}}  \nonumber \\
 &&\times \Psi_{_{^3\text{He}}}^{^{\lambda_{^3\text{He}}}}(\beta_1=1/3,\lambda_1,p_{_{1\perp}},\beta_2,p_{_{2\perp}}\lambda_2,\lambda_3) 
  \frac{d^2p_{d\perp}}{(2\pi)^2} 
\frac{d \beta_{3}}{\beta_3} \frac{d^2p_{3\perp}}{2(2\pi)^3}\frac{d \alpha_{3}'}{\alpha_{3}'} \frac{d^2p_{3\perp}'}{2(2\pi)^3}  \nonumber \\
&&  \times \frac{dx_1}{x_1}\frac{d^2k_{1\perp}}{2(2\pi)^3} \frac{dx_{2}'}{x_{2}'}\frac{d^2k_{2\perp}}{2(2\pi)^3}.
\end{eqnarray}
%The charge $Q_i$ of the struck quark needs to be summed over to account for all possible photon quark interactions. In the approximation of $\beta_1 = 1/3$, above equation is the most general expression for the scattering amplitude for the reaction $\gamma + ^3\text{He} \rightarrow p + d$.

\appendix
\renewcommand{\theequation}{B-\arabic{equation}}
\renewcommand\thefigure{B.\arabic{figure}}
\setcounter{figure}{0}  
\setcounter{equation}{0}
\section*{APPENDIX B:\quad High momentum transfer pd $\rightarrow$ pd scattering}
\begin{figure}[ht]
\centering
 \includegraphics[width=0.8\textwidth]{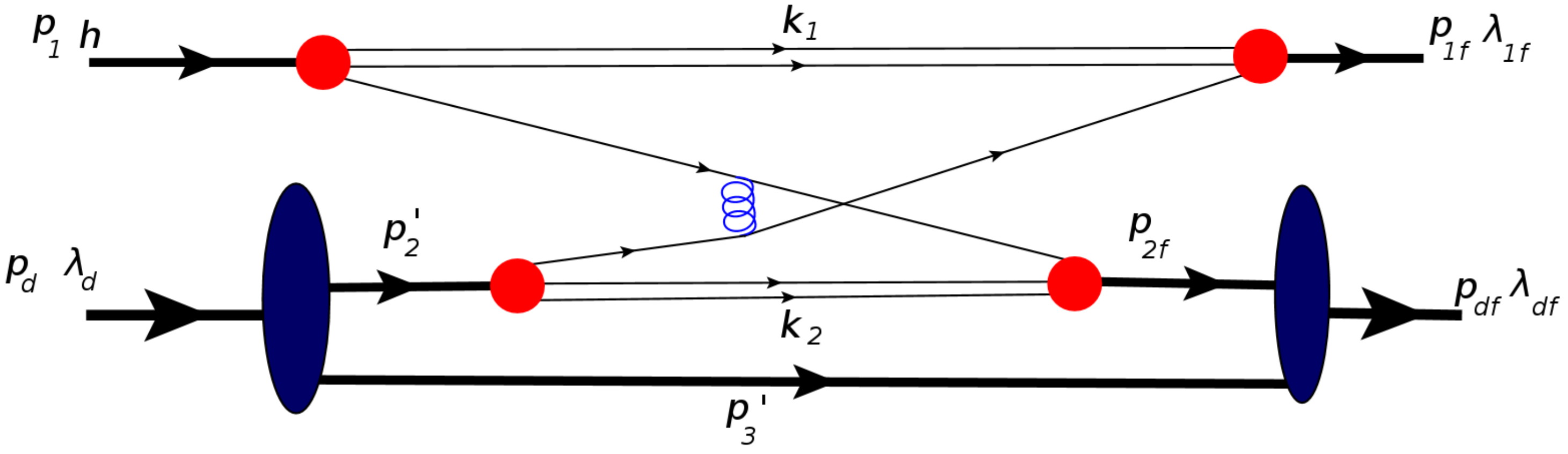}
\caption{Typical quark-interchange mechanism of hard $pd \rightarrow pd$ scattering.}
\label{Fig_B1}
\end{figure}

In this section, we study the high momentum transfer elastic proton - deuteron scattering
based on the quark-interchange mechanism.  A characteristic diagram of such scattering is shown in Fig.\ref{Fig_B1}. 
The notations in this figure are chosen to be similar to the $pd\rightarrow pd$ rescattering part of the $\gamma ^3\text{He}\rightarrow pd$ amplitude in Eq. (\ref{AmpToIntroducePDamp}).
%Here, the initial and final momenta of the proton are given by $p_1$ and $p_{1f}$ whereas the initial and final momenta of the deuteron are 
%$p_{d} = p_{2}' + p_{3}'$ and $p_{df}$ respectively. 
Here the helicities in the initial and final states of the proton are $h$ and $\lambda_{1f}$ and for the 
deuteron  they are $\lambda_{d} $ and $\lambda_{df}$. The momenta defined in Fig. \ref{Fig_B1} satisfy the following 
four-momentum conservation relations:
\begin{equation}
p_{1} + p_{d} = p_{1f} + p_{df}, \qquad p_{d} = p_{2}' + p_{3}'.
\end{equation}

The Feynman amplitude for this $ pd \rightarrow pd $ scattering can be written as follows:
\begin{eqnarray}
\mathcal{M}_{pd}= && \nonumber \\
N1: &&\int \bar{u}(p_{1f}, \lambda_{1f}) (-i) \Gamma_{n1}^{\dagger} \frac{i(\slashed{p}_{1f} -\slashed{k}_{1} + m_q)}{(p_{1f} - k_{1})^2 -m_q^2 + i\epsilon} \frac{iS(k_1)}{k_1^2 - m_s^2 + i\epsilon} [-igT_c^\beta \gamma_\mu]  \nonumber \\
&&\times \frac{i(\slashed{p_{1}} -\slashed{k_{1}} + m_q)}{(p_{1} - k_{1})^2 -m_q^2 + i\epsilon} i\Gamma_{n1} u(p_1, \lambda_1) \frac{d^4k_1}{(2\pi)^4} \nonumber \\
D-N2: && \int \chi_{df}^\dagger (-i)\Gamma_{DNN}^\dagger \frac{i(\slashed{p}_{2f} + m)}{p_{2f}^2 - m_N^2 + i\epsilon} \bar{u}(p_{2f},\lambda_{2f}) (-i) \Gamma_{n2f}^\dagger \frac{i(\slashed{p}_{2f} -\slashed{k}_{2} + m_q)}{(p_{2f} - k_{2})^2 -m_q^2 + i\epsilon}  \nonumber \\
 && \times \frac{i(\slashed{p_3}'+m)}{p_3'^{2}-m_N^2 + i\epsilon} 
 \frac{iS(k_2)}{k_2^2 - m_s^2 + i\epsilon} [-igT_c^\alpha \gamma_\nu] \frac{i(\slashed{p_{2}}' -\slashed{k_{2}} + m_q)}{(p_{2}' - k_{2})^2 -m_q^2 + i\epsilon} i\Gamma_{n2'} u(p_{2}' ,\lambda_{2}')  \nonumber \\
&& \times \frac{i(\slashed{p}_{2}' + m)}{p_{2}'^2 - m_N^2 + i\epsilon} 
  i\Gamma_{DNN} \chi_d \frac{d^4k_2}{(2\pi)^4} \frac{d^4p_{3}'}{(2\pi)^4} \nonumber \\
g: && \frac{i d_{\mu \nu} \delta_{\alpha \beta}}{q_{q}^{2}}.
\end{eqnarray}
The following derivations are analogous to that of Eq. (\ref{StartDerivationAmplitude}), where we identify the parts associated with the deuteron wave function 
as well as with the quark wave functions of nucleons and perform integrations corresponding to the on-shell conditions for the spectator nucleon in the deuteron and 
spectator quark-gluon states in the nucleons.
We first consider the expression for ${\bf N1}$, for  which, performing  derivations similar to those for the $N1$ term in  Eq. (\ref{StartDerivationAmplitude}) and using 
the definition of the quark wave function of the nucleon according to  Eq. (\ref{QuarkWF}) one obtains
\begin{eqnarray} \label{N1_B1_with_QWF}
N1:= \sum_{\substack{\lambda_1 \\ \eta_{1}, \eta_{1f} }} 
&& -i \int	\frac{\Psi_{n1f}^{\dagger \lambda_{1f};\eta_{1f}} (x_{s1},k_{1\perp}, p_{1f\perp})}{1 - x_{s1}} 
\bar{u}_q(p_{1f} - k_1,\eta_{1f}) u_q(p_1 - k_1,\eta_1)\Big[ -igT_{c}^\beta \gamma_{\mu}\Big]  \nonumber \\ 
&& \times \frac{\Psi_{n1}^{\lambda_1;\eta_1}(x_1, k_{1\perp},p_{1\perp}) }{1 - x_1} \frac{1}{2}\dfrac{dx_1}{x_1} \dfrac{d^2k_{1\perp}}{(2\pi)^3}. 
\end{eqnarray}
With similar derivations in the ${\bf D-N2}$ part and using, in addition to the quark wave function of nucleon, the deuteron light-front wave function defined 
in  Eq. (\ref{DeuteronWF}) one obtains
\begin{eqnarray} \label{B_N2final}
D-N2 = && \sum_{\substack{ \lambda_{d},\lambda_{2f}, \lambda_{2}' ,\lambda_{3}' \\ \eta_{2}', \eta_{2f} }}  -i \int \frac{\Psi_{df}^{\dagger \lambda_{df}:\lambda_{3}', \lambda_{2f}} (\alpha_3'/\gamma_{d}, p_{df\perp},p_{3\perp}')}{1-\alpha_3'/\gamma_{d}} \bar{u}_q(p_{2f} - k_2,\eta_{2f})   \nonumber \\
&& \times \frac{\Psi^{\dagger \lambda_{2f};\eta_{2f}}_{n2f}(x_{s2},k_{2\perp},p_{2f\perp})}{1-x_{s2}} [ -igT_c^\alpha \gamma_\nu] 
 \frac{\Psi_{d}^{\lambda_{d}:\lambda_{2}', \lambda_{3}'}(\alpha_{3}',p_{d\perp},p_{3\perp}')}{1-\alpha_{3}'} u_q(p_2'-k_2,\eta_{2}')    \nonumber \\
 &&\times \frac{\Psi_{n2'}^{\lambda_{2}';\eta_{2}'}(x_{2}',k_{2\perp},p_{2\perp}')}{1-x_{2}'} \frac{1}{2} \frac{dx_{2}'}{x_{2}'} \frac{d^2k_{2\perp}}{(2\pi)^3} \frac{d\alpha_{3}'}{\alpha_{3}'} \frac{d^2p_{3\perp}'}{2(2\pi)^3}.
\end{eqnarray} 
 
Combining   Eqs.(\ref{N1_B1_with_QWF}) and (\ref{B_N2final})  for the amplitude of $pd\rightarrow pd$ scattering, one arrives at
\begin{eqnarray} \label{pd_amplitude}
&&\mathcal{M}_{pd}^{\lambda_{df},\lambda_{1f};\lambda_{d},\lambda_1}  = \sum_{\substack{(\lambda_{2f})(\lambda_1, \lambda_d)(\lambda_{2}',\lambda_{3}')\\ (\eta_{1f}, \eta_{2f})(\eta_{1}, \eta_{2}') }}
 \int	\frac{\Psi_{df}^{\dagger \lambda_{df}: \lambda_{3}',\lambda_{2f}} (\alpha_3'/\gamma_{d}, p_{df\perp},p_{3\perp}')}{1-\alpha_3'/\gamma_{d}} \Bigg\{\frac{\Psi^{\dagger \lambda_{2f};\eta_{2f}}_{n2f}(x_{s2},k_{2\perp},p_{2f\perp})}{1-x_{s2}}  \nonumber \\
 && \times \bar{u}_q(p_{2f} - k_2,\eta_{2f})
 [ -igT_c^\alpha \gamma_\nu]
  u_q(p_1 - k_1,\eta_1) 
 \frac{\Psi_{n1}^{\lambda_1;\eta_1}(x_1, k_{1\perp},p_{1\perp}) }{1 - x_1}\Bigg\}_{1} G^{\mu \nu}(r) \nonumber \\
&& \times \Bigg\{\frac{\Psi_{n1f}^{\dagger \lambda_{1f};\eta_{1f}} (x_{s1},k_{1\perp}, p_{1f\perp})}{1 - x_{s1}}
 \bar{u}_q(p_{1f} - k_1,\eta_{1f})
 [ -igT_c^\beta \gamma_\mu] u_q(p_2'-k_2,\eta_{2}')  
 \frac{\Psi_{n2'}^{\lambda_{2}';\eta_{2}'}(x_{2}',k_{2\perp},p_{2\perp}')}{1-x_{2}'} \Bigg\}_{2}  \nonumber \\
  && \times \frac{\Psi_{d}^{\lambda_d: \lambda_{2}', \lambda_{3}'}(\alpha_{3}',p_{d\perp},p_{3\perp}')}{1- \alpha_{3}'} 
   \frac{1}{2}\dfrac{dx_1}{x_1} \dfrac{d^2k_{1\perp}}{(2\pi)^3} \frac{1}{2} \frac{dx_{2}'}{x_{2}'} \frac{d^2k_{2\perp}}{(2\pi)^3} \frac{d\alpha_{3}'}{\alpha_{3}'} \frac{d^2p_{3\perp}'}{2(3\pi)^3}.
\end{eqnarray}

\appendix
\renewcommand{\theequation}{C-\arabic{equation}}
\renewcommand\thefigure{C.\arabic{figure}}
\setcounter{figure}{0}  
\setcounter{equation}{0}
\section*{APPENDIX C: Relating the Light-Front and Non-Relativistic Wave Functions}
To obtain the relation between light-front and  nonrelativistic nuclear wave functions in the small-momentum limit, we  consider the fact that the 
light-front nuclear wave function is normalized based on baryonic number conservation~(see, e.g., Refs.\cite{Frankfurt:1985ui,srcrev,Cosyn:2010ux}),  
while the nonrelativistic (Schroedinger)  wave function is normalized as $\int |\Psi_A(p)|^2 d^3p = 1$.

\begin{figure}[ht]
\centering
\includegraphics[scale=0.3]{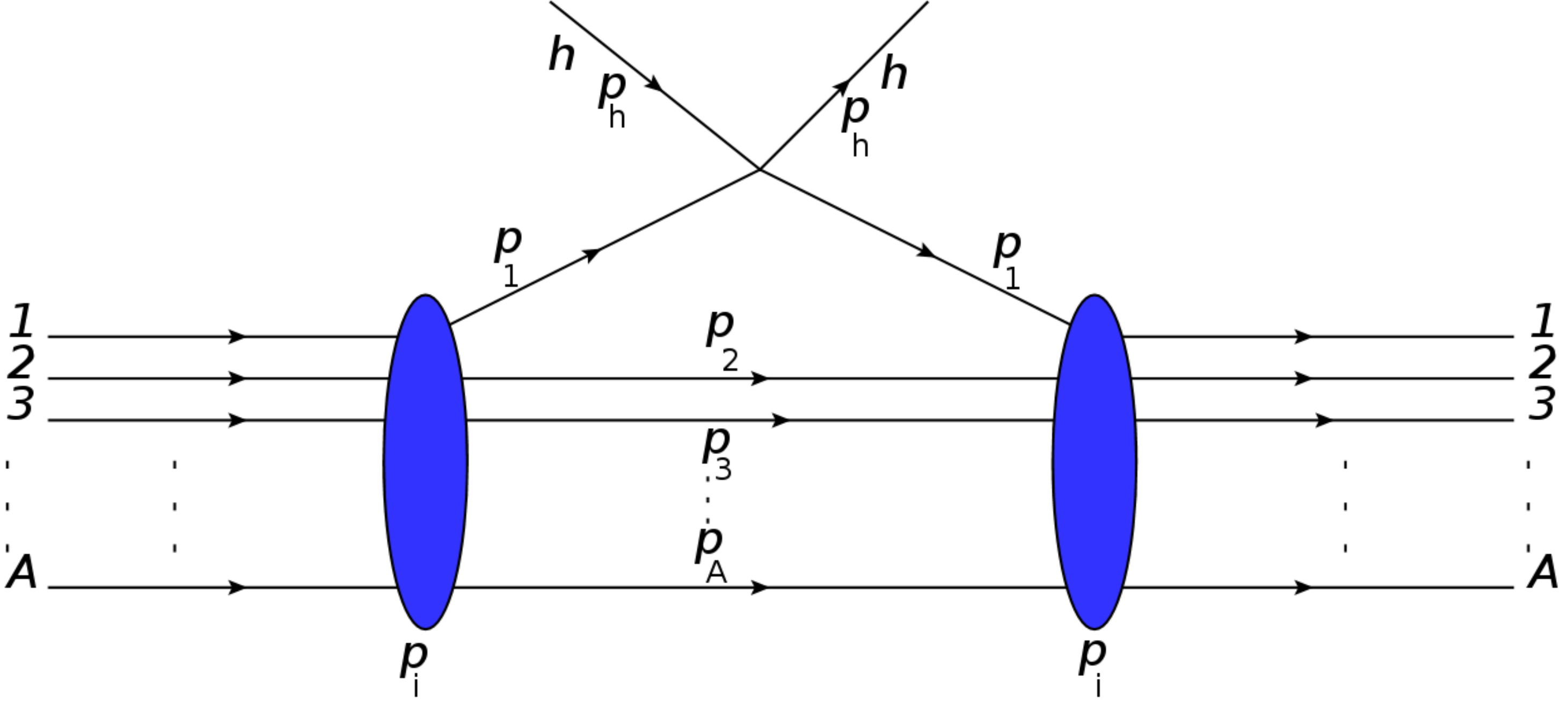}
\caption{Hadronic probe to see baryons in a nucleus.}
\label{probe_diagram}
\end{figure}

To obtain the normalization condition based on baryonic number conservation, we consider  a $h+A\rightarrow h+A$ scattering in forward 
direction  in which $h$ probes the 
constituent baryons in the nucleus $A$ (see Fig. \ref{probe_diagram}).  In the figure we assign $p_i$ to be the four-momentum of the nucleus while
$p_1, p_2, \cdots, p_A$ are four-momenta of constituent nucleons such that $p_1 + p_2 + \cdots + p_A = p_i$. 
For the diagram of Fig. \ref{probe_diagram}, applying the Feynman rules one obtains
\begin{eqnarray} \label{hadron_probe_amplitude}
\mathcal{M}_{hA} &&= \sum_{N} \int \chi_A^{\dagger} \Gamma_A^{\dagger} \frac{\slashed{p}_1 + m}{p_1^2 - m_N^2 + i\epsilon}
 \hat{M}_{hN} \frac{\slashed{p}_A + m}{p_A^2 - m_N^2 + i\epsilon} \cdots \frac{\slashed{p}_2 + m}{p_2^2 - m_N^2 + i\epsilon} 
 \frac{\slashed{p}_1 + m}{p_1^2 - m_N^2 + i\epsilon} \Gamma_A  \chi_A  \nonumber \\
&& \times \frac{d^4p_2}{(2\pi)^4} \frac{d^4p_3}{(2\pi)^4} \cdots \frac{d^4p_A}{(2\pi)^4},
\end{eqnarray}
where we sum over all the possible nucleons that can be probed and ${\hat M}_{hN}$ represents the effective vertex of the hadron-nucleon interaction.
We use the sum rule for the spinors and also integrate by the minus component of the momenta using the scheme given in 
Eq. (\ref{polevalueint_scheme}), to obtain
\begin{eqnarray}
  \mathcal{M}_{hA} & = &   \sum_{N} \sum_{\lambda_1, \lambda_2, \cdots \lambda_A} \int \chi_A^{\dagger} \Gamma_X^{\dagger} \frac{u(p_1, \lambda_1) \bar{u}(p_1, \lambda_1) }{p_{1+} \big(p_{1-} - \frac{m_N^2 + p_{1\perp}^2}{p_{1+}}\big)} \hat{M}_{hN} u(p_A , \lambda_A) \bar{u}(p_A, \lambda_A)\cdots   \nonumber \\ 
&& \times  u(p_3 , \lambda_3) \bar{u}(p_3, \lambda_3) u(p_2 , \lambda_2) \bar{u}(p_2, \lambda_2) \frac{u(p_1 , \lambda_1) \bar{u}(p_1, \lambda_1)}{p_{1+} \big(p_{1-} - \frac{m_N^2 + p_{1\perp}^2}{p_{1+}}\big)}  \Gamma_A  \chi_A  \nonumber \\
& & \times \frac{dp_{2+}}{p_{2+}} \frac{d^2p_{2\perp}}{2(2\pi)^3} \frac{dp_{3+}}{p_{3+}} \frac{d^2p_{3\perp}}{2(2\pi)^3} \cdots \frac{dp_{A+}}{p_{A+}} \frac{d^2p_{A\perp}}{2(2\pi)^3}, 
\end{eqnarray}
where the $\lambda_j$ denote the helicities of the nucleon with momentum $p_j$. Considering the transverse momentum of the nucleus $A$ to be zero, we note that
\begin{eqnarray} \label{P1minus_Xnuc}
&&p_{1-} - \frac{m_N^2 + p_{1\perp}^2}{p_{1+}} = p_{i-} - p_{2-} - p_{3-}- \cdots - p_{A-} - \frac{m_N^2 + p_{1\perp}^2}{p_{1+}}  \nonumber \\
%&& = \frac{m_N^2}{p_{i+}} - \frac{m_N^2 + p_{2\perp}^2}{p_{2+}} - \frac{m_N^2 + p_{3\perp}^2}{p_{3+}} - \cdots - \frac{m_N^2 + p_{A\perp}^2}{p_{A+}} - \frac{m_N^2 + p_{1\perp}^2}{p_{1+}} \nonumber \\
&&= \frac{1}{p_{i+}}\Big[ m_N^2 - \frac{m_N^2 + p_{2\perp}^2}{\beta_2} - \frac{m_N^2 + p_{3\perp}^2}{\beta_3} - \cdots - \frac{m_N^2 + p_{A\perp}^2}{\beta_A } - \frac{m_N^2 + p_{1\perp}^2}{\beta_1}\Big],
\end{eqnarray}
where  $\beta_j = \frac{p_{j+}}{p_{i+}}$ are the light-front momentum fractions of the nucleus $A$ carried by the nucleons $j$ ($j = 1,\cdots A$).
Introducing the Feynman amplitude for $h+N\rightarrow hN$ as $\mathcal{M}_{hN} = \bar{u}(p_1, \lambda_1) \hat{M}_{hN}  u(p_1, \lambda_1)$ for Eq. (\ref{P1minus_Xnuc})
one obtains
\begin{eqnarray} \label{Xnuc_amplitude}
& \mathcal{M}_{hA} &= \sum_{N}\sum_{\lambda_1, \lambda_2, \cdots \lambda_A} \int \chi_A^{\dagger} \Gamma_A^{\dagger} \frac{u(p_1, \lambda_1) u(p_2, \lambda_2)u(p_3, \lambda_3) \cdots u(p_A, \lambda_A) }{\frac{p_{1+}}{p_{i+}} \Big[ m_N^2 - \frac{m_N^2 + p_{2\perp}^2}{\beta_2} - \frac{m_N^2 + p_{3\perp}^2}{\beta_3} -\cdots -  \frac{m_N^2 + p_{A\perp}^2}{\beta_A} \Big]} \mathcal{M}_{hN}  \nonumber \\
&& \times \frac{ \bar{u}(p_1, \lambda_1) \bar{u}(p_2, \lambda_2)  \bar{u}(p_A, \lambda_A)}{\frac{p_{1+}}{p_{i+}} \Big[ m_N^2 - \frac{m_N^2 + p_{2\perp}^2}{\beta_2} - \frac{m_N^2 + p_{3\perp}^2}{\beta_3} -\cdots -  \frac{m_N^2 + p_{A\perp}^2}{\beta_A} \Big]} \Gamma_A \chi_A \prod_{k = 2}^{A} \frac{d\beta_{k}}{\beta_k} \frac{d^2p_{k\perp}}{2(2\pi)^3}.
\end{eqnarray}
Using the generalization of Eqs.(\ref{He3wf}) and (\ref{DeuteronWF}) for light-front nuclear wave function of nucleus $A$, the above equation reduces to
\begin{eqnarray} \label{Xnuc_amp_with_wfs}
\mathcal{M}_{hA} &&= \sum_{N} \sum_{\lambda_1, \lambda_2, ... \lambda_A} \int \frac{\Psi_A^{LC \dagger}( \beta_2, \beta_3, ... \beta_A, p_{2\perp}, p_{3\perp} ... p_{A\perp}, \lambda_2, \lambda_3, ... \lambda_A) }{\beta_1} \mathcal{M}_{hN}  \nonumber \\
&&\times \frac{\Psi_A^{LC}( \beta_2, \beta_3, ... \beta_A, p_{2\perp}, p_{3\perp} ... p_{A\perp}, \lambda_2, \lambda_3, ... \lambda_A )}{\beta_1} \prod_{k = 2}^{A} \frac{d\beta_{k}}{\beta_k} \frac{d^2p_{k\perp}}{2(2\pi)^3}.
\end{eqnarray}

We now make use of the optical theorem,   according to which  
%allows us to relate the imaginary part of the forward scattering amplitude to the 
%total cross section in high energy limit as follows:
\begin{equation}
\text{Im } \mathcal{M}_{hA} = s_{hA} \sigma_{hA} \text{ and } \text{Im } \mathcal{M}_{hN} = s_{hN} \sigma_{hN}, 
\label{OpTheorem}
\end{equation}
where $s_{hA}= (p_h + p_i)^2$   and 
$\sigma_{hA}$ is the total cross section of $hA$ scattering.  Similarly,  $s_{hN}$ and $\sigma_{hN}$ are invariant energy 
and  total cross section for $hN$ scattering.
 The conservation of baryon number allows us to relate $\sigma_{hA} = A\sigma_{hN}$. Using this relation together with Eq. (\ref{OpTheorem})  in
 Eq. (\ref{Xnuc_amp_with_wfs}), one obtains
\begin{equation} \label{lightcone_norm}
\int \frac{\left| \Psi_A^{LC}( \beta_2, \beta_3, ... \beta_A, p_{2\perp}, p_{3\perp} ... p_{A\perp}, \lambda_2, \lambda_3, ... \lambda_A) \right|^2}{\beta_1^2} \frac{s_{hN}}{s_{hA}} \prod_{k = 2}^{A} \frac{d\beta_k}{\beta_k} \frac{d^2p_{k\perp}}{2(2\pi)^3} = 1.
\end{equation}
To obtain the relation of light-front wave function to the nonrelativistic wave function in the small-momentum limit, we note that in such limit
$\beta_k = \frac{E_k + p_k^z}{p_{i+}}\approx 1+ \frac{p_{k}^z}{ m_N}$ thus $\frac{d\beta_k}{\beta_k} = \frac{dp_k^z}{m_N}$. 
Furthermore, in the high energy limit of the hadronic probe in which large momentum of the hadrons points in the $-\hat z$ direction, 
$s_{hA} \approx p_{h-}p_{A+}$ and  $s_{hN} \approx p_{h-}p_{N+}$ resulting in
\begin{equation}
\frac{s_{hN}}{s_{hA}} = \frac{p_{N+}}{p_{A+}} = \frac{\beta_1}{ A}.
\end{equation}
Applying all these approximations in Eq. (\ref{lightcone_norm})  one obtains
\begin{equation} \label{normalized_eqn_a}
\int \frac{\left| \Psi_A^{LC}( \beta_2, \beta_3, ... \beta_A, p_{2\perp}, p_{3\perp} ... p_{A\perp}, \lambda_2, \lambda_3, ... \lambda_A) \right|^2} {1/A}   \frac{1}{m_N^{A-1} [2(2\pi)^3]^{A-1}} \prod_{k = 2}^{A}d^3p_k = 1.
\end{equation}
Next we compare the above expression with the normalization condition for the nonrelativistic Schroedinger wave function:
\begin{equation} \label{normalized_eqn_b}
\int \left| \Psi_A^{NR}(\vec{p}_1, \vec{p}_2, ... \vec{p}_A )\right|^2 \prod_{k = 2}^{A} d^3p_k = 1,
\end{equation}
where  $\vec p_1 = \vec p_i -  \vec p_2 - \cdots -  \vec p_A$. This comparison allows us to relate the light-front nuclear wave function and 
the Schroedinger wave function in the following form:
\begin{equation} \label{LCNR_relation_final}
\Psi_X^{LC} ( \beta_1, \beta_2, ..., p_{1\perp}, p_{2\perp}...) = \frac{1}{\sqrt{A}} \big[m_N 2(2\pi)^3\big]^{\frac{A-1}{2}} \Psi_X^{NR}(\vec{p}_1, \vec{p}_2, ... ).
\end{equation}

\end{document}